# A binary-related origin mediated by environmental conditions for blue straggler stars

Francesco R. Ferraro[1,2*], Barbara Lanzoni[1,2], Enrico Vesperini[3], Emanuele Dalessandro[2], Mario Cadelano[1,2], Cristina Pallanca[1,2], Giacomo Beccari[4], Domenico Nardiello[5,6], Mattia Libralato[6], and Giampaolo Piotto[5]

[1]*Dipartimento di Fisica e Astronomia "Augusto Righi", Alma Mater Studiorum Universita` di Bologna, Via Piero Gobetti 93/2, I-40129 Bologna, Italy*

[2]*INAF -- Astrophysics and Space Science Observatory Bologna, Via Piero Gobetti 93/3, I-40129 Bologna, Italy*

[3] *Department of Astronomy, Indiana University, Bloomington, IN, 47401, USA*

[4]*European Southern Observatory, Karl-Schwarzschild-Strasse 2, 85748 Garching bei Munchen, Germany*

[5]*Dept. of Physics and Astronomy Galileo Galilei, University of Padova, Vicolo dell'Osservatorio 3, I-35122 Padova, Italy*

[6]*Istituto Nazionale di Astrofisica - Osservatorio Astronomico di Padova, Vicolo dell'Osservatorio 5, I-35122, Padova, Italy*

**Blue stragglers are anomalously massive core hydrogen-burning stars that, according to the theory of single star evolution, should not exist. They are suspected to form in mass-enhancement processes, involving binary evolution or stellar collisions. In dynamically active systems like globular clusters, the number of blue stragglers originated by collisions is expected to increase with the local density and the rate of stellar encounters. Here we analyse more than 3000 blue stragglers in 48 Galactic globular clusters with different structures, finding that their number normalized to the sampled luminosity anti-correlates (instead of correlating) with the central density, collision rate, and dynamical age of the parent cluster. Similar trends are also found for the cluster binary fraction. Once inserted in the context of the current knowledge of the BSS phenomenon, these correlations indicate that low-density regions (possibly because of a higher binary production/survival rate) are the natural habitat of both BSSs and binary systems, and the observed BSSs mostly have a binary-related origin mediated by the environmental conditions.**

## Introduction

Despite their mysterious origin, blue straggler stars (BSSs) are routinely observed in all resolved stellar systems[1-8]. The current scenarios suggest that BSSs are generated from mass-transfer activity[9] in binary systems, direct stellar collisions[10-12], or mergers in binary or higher order systems[13], possibly mediated by dynamical interactions[14,15]. In this framework, Galactic globular clusters (GGCs) are ideal cosmic laboratories to study BSSs, since they host a non-negligible fraction of binaries[16] and they also are collisional stellar systems, where frequent gravitational interactions among stars occur on timescales shorter than their age[17], thus possibly harbouring a significant population of collisional BSSs[10-12,14,15,18]. In addition, the efficiency of the different formation channels is expected to be modulated by the environmental conditions, and GGCs span large ranges of physical properties[19] such as total mass, central density, relaxation time, and probability of gravitational interactions. Numerical simulations[14,15] predict that binary-mediated collisions should be dominant in GGCs, and a possibly weak, but positive correlation between the BSS number and the collision rate should be present. All this has made the search for possible links between the BSS content in GGCs and the properties of the host system one of the primary goals in BSS studies. Despite intense efforts, however, the only two clean relations found so far are with the total luminosity of the parent cluster[3,20-22] and between the number of core BSSs and the core mass[23], while controversial or poorly significant results have been obtained for the other cluster properties (see Discussion).

Here we present the analysis of a sample of 3419 BSSs homogeneously observed and selected in 48 GGCs with different structural parameters. We find clear-cut relations with the parent cluster environmental conditions demonstrating that the number of BSSs per unit luminosity (the BSS specific frequency) significantly decreases (by more than one order of magnitude) with the cluster total luminosity, central velocity dispersion, central surface brightness, central density, collisional parameter, and dynamical age. We determine impressively similar relations also for the overall binary fraction ($f_{bin}$) and demonstrate that these, together with the tight correlation found between the BSS specific frequency and $f_{bin}$, are able to fully reproduce the trends observed for BSSs. The collected evidence leads to the conclusion that BSSs are predominantly formed from binary systems (even in high-density clusters) and the overall binary (and binary by-product) content is set by the environmental conditions, with a potential impact in a broad range of astrophysical research areas.

## Results

### Defining the BSS sample

The present study takes advantage of a sample of 48 GGCs observed in a homogeneous way[24] through high-resolution images collected with Wide Field Camera 3 (WFC3) onboard the Hubble Space

Telescope (HST) in the F275W and F336W filters. The dataset and the data reduction procedures are described in previous papers[25,26] and are summarized in section "Data analysis" in Methods. The ($m_{F275W}$, $m_{F275W}$-$m_{F336W}$) colour-magnitude diagram (CMD) is optimal for the BSS selection[2,25], since these stars are well distinguishable from the other evolutionary sequences, defining a clear-cut sequence populating an almost diagonal strip that has a vertical extension of about 3 magnitudes and it is approximately 2 magnitudes wide in colour. To perform a homogeneous selection of BSSs in clusters with different values of distance, reddening and metallicity, we followed the procedure adopted in previous papers[25,27]. After correction for differential reddening and membership selection through relative proper motions[28] (see "Differential reddening correction and proper motion selection" in Methods), we defined a normalized CMD (n-CMD) where all the stellar magnitudes and colours measured in a given cluster are arbitrarily shifted to locate the main sequence turnoff (MSTO) at magnitude $m^*_{F275W}$=0 and colour $(m_{F275W}-m_{F336W})^*$=0 (see the details in the Section "Constructing the normalized CMD" in Methods). Thus, all the BSSs populate the same region in the n-CDM, and their values of $m^*_{F275W}$ and $(m_{F275W}-m_{F336W})^*$ directly quantify how much they are brighter and bluer than the cluster MSTO stars. We performed a homogeneous selection of the BSSs belonging to each cluster by adopting a unique box in the n-CMD, with the boundaries defined in previous studies[25] to include the bulk of the populations. Figure 1a shows the n-CMD of the total sample of proper motion selected BSSs detected in the 48 considered clusters, counting a grand-total of 3419 objects. The distribution of the number of BSSs observed in each cluster ranges from a minimum of 12 BSSs detected in NGC 6535, to 179 BSSs found in NGC 5286, with a peak around 50-60, a pronounced tail toward larger values, and an average value of $N_{BSS}$=71 ±39.

**The BSS specific frequency**

The sample refers to the central 85" region of the selected clusters, corresponding to the field of view covered by the observations. Of course, this corresponds to different amounts of sampled luminosity ($L_{samp}$) because the 48 GGCs are characterized by different distances from Earth (ranging between 2 and 18 kpc) and different intrinsic sizes and structures, with structural parameters that cover the entire range of values spanned by the Galactic population in terms of total luminosity, central density, core radius, etc. (see Supplementary Information and Supplementary Data 1). This implies that the number of "normal" stars, like horizontal branch (HB), Red Giant Branch (RGB) or MSTO stars, changes from one cluster to another proportionally to the amount of luminosity sampled by the observations. Hence, a meaningful comparison of the content of normal stars among the different clusters would require the normalization of their observed numbers to $L_{samp}$: indeed, the specific frequency (i.e., the number of stars divided by $L_{samp}$) of any post-MS population is set by the duration of the corresponding evolutionary phase[29], which is the same in clusters of any mass. On the other hand, BSSs are an exotic population that not necessarily follows the same rules. Indeed, Figure 1b clearly shows that also the number of detected BSSs positively scales with $L_{samp}$ (as determined by integrating the King model that best-fits the observed density profile; see "Sampled luminosity and sampled populations" in Methods), although with a shallower slope (just 0.4). From one side, this evidence suggests that to properly compare the BSS content in intrinsically different systems, we need to normalize the BSS number to $L_{samp}$ (see also Discussion). Hence, here we compute the BSS specific frequency $F4_{BSS}$, defined as the number of identified BSSs normalized to the sampled luminosity in units of $10^4\ L_\odot$. This quantity was introduced in the very first quantitative studies[3,30] of BSSs in GGCs, and it provides the number of BSSs that any given environment hosts per unit luminosity or per unit mass, corresponding to the BSS production/survival efficiency. Among the considered clusters, we find that $F4_{BSS}$ varies by a factor of 20, ranging from 3 to 58 BSSs for each $10^4\ L_\odot$ sampled luminosities, with a peak at approximately 8 and an average value of $14.5 \pm 11.3$. (see Fig. 1c). This means that the BSS production/survival efficiency can be very different in different clusters. From the other side, the shallow dependence of $F4_{BSS}$ on the sampled luminosity testifies another key difference with respect to "normal" stellar populations: while the number of standard stars normalized to the sampled luminosity is essentially constant (being only set by the duration of the specific evolutionary stage), the BSS specific frequency still depends on $L_{samp}$, indicating that additional processes are playing a role in determining it. Since the sampled luminosity depends on cluster parameters such as the total luminosity or the central density, this suggests that also the BSS production/survival efficiency should depend on these environmental properties.

**Correlations between BSS frequency and environment**

In agreement with previous studies[3,20-23] (see Discussion for a summary of previous results in this respect), we find that the BSS specific frequency decreases as function of the total luminosity of the parent cluster (Fig. 2a; also see Supplementary Table 1 for the values of the best-fit parameters and the Pearson correlation coefficient). In addition, here we show a clear anti-correlation with the central velocity dispersion $\sigma_0$ (Fig.2b). This was only barely seen in previous works[20,21], although it was expected since $\sigma_0$ strongly correlates with the cluster mass (hence, luminosity).

However, the most intriguing result comes from the comparison between the BSS specific frequency and other properties characterizing the cluster environment, such as the central density and the collision rate. In fact, while high-density high-collisional environments are expected to favour the activation of the BSS collisional channels[14,15], we find instead that the BSS specific frequency decreases in these conditions. This is illustrated in Figure 2c,d,e where the measured BSS specific frequency is plotted as a function of the central surface brightness[19] corrected for interstellar extinction ($\mu_{V,0}$), the central luminosity density ($\log \rho_0$)[19], and the collisional parameter[31] $\Gamma_{S+S}$ of the parent cluster (see Section "The single-single collisional parameter" in Methods for the description of how this parameter has been estimated). The values of the Pearson correlation coefficient (see Supplementary Table 1) indicate statistically robust linear dependences (so far, no or just weak anti-correlations had been found[3,20,21,23,32,33] with these quantities: see Discussion). It is worth noticing that no significant trends with these parameters are observed for normal stellar populations, indicating that they are peculiar to BSSs. We also emphasize that the detected trends are confirmed (although less pronounced and less significant) even if we consider just the BSS population within the cluster core or in a fraction of the half-mass radius (see Supplementary Fig. 1). Moreover, the detected anti-correlations turn out to be independent of the normalization quantity adopted in the definition of $F4_{BSS}$. In fact, the same trends remain even if $L_{samp}$ is computed as the sum of the luminosity of all the detected stars (see Supplementary Fig. 2), or the normalization is done with respect to a reference population of normal cluster stars, such as the HB or MSTO and sub-giant branch populations (see Supplementary Fig.3).

Since the internal dynamical evolution tends to progressively increase the central density (and thus the collisionality) of star clusters, a dependence of the BSS frequency is expected also as a function of the host system dynamical age, and it is indeed observed. As a measure of cluster dynamical ages, here we adopted the $A^+$ parameter[34,35], quantifying the level of BSS central segregation due to the action of dynamical friction, which makes the most massive objects (such as BSSs) sink to the cluster centre as function of the dynamical ageing of the system[25,36-39] (see Section "Measuring dynamical evolution with the BSS dynamical clock" in Methods), with larger values of $A^+$ corresponding to dynamically older clusters[25]. The BSS specific frequency is found to decrease with $A^+$ (see Figure 2f; see also Supplementary Figure 4 for the trend with the central relaxation time[19]), in agreement with the expected effect of dynamical ageing and nicely supporting the emerging scenario. We found no correlation with any other cluster parameter (such as metallicity, age, distance, reddening).

These observational findings clearly demonstrate that, at odds with the case of normal stars, the number of BSSs per unit luminosity (or per unit mass) varies significantly depending on the cluster total mass, central density, collisional parameter, and dynamical age. The evidence that the same amount of cluster luminosity (mass) in systems with different structural parameters hosts a different number of BSSs points toward the occurrence of specific processes that tend to favour or disfavour the formation or the survival of BSSs depending on the local environment. More specifically, in high density conditions, the efficiency of BSS formation/survival is up to 20 times lower than in "peaceful", low-density environments.

**Correlations between binary fraction and environment**

As matter of fact, the most efficient BSS formation channels proposed so far are all directly related to binaries. In fact, besides the case of secular evolution of dynamically unperturbed primordial binaries, numerical simulations[14,15] show that dynamical encounters involving binaries (binary-single and binary-binary interactions) are by far the most effective collisional channels, with respect to single-single encounters. Binary-binary and single-binary interactions should also have a role in hardening primordial binaries[12,18,14,15] (therefore promoting the formation of mass-transfer BSSs) and may induce stellar mergers in binary or even triple systems[13]. Hence, we extended our investigation to the cluster binary fraction ($f_{bin}$). To this purpose, we took advantage of the photometric estimate of $f_{bin}$ homogeneously determined[16] for a large number of GGCs (see "The binary fraction" in Methods). Very interestingly, we find that the overall binary fraction of the parent cluster shows the same correlations and anti-correlations (see Figure 3) discovered for the BSS specific frequency: $f_{bin}$ decreases for increasing cluster mass, central density, collisional parameter and dynamical age of the parent cluster. The observed best-fit slopes are impressively similar to those found for BSSs (see Supplementary Table 1).

**BSSs: a binary origin mediated by environmental conditions**

These findings not only show that binaries preferably form/survive in low-density environments, but also suggest that the overall amount of binary systems hosted in the parent cluster could be the key parameter responsible for the observed BSS populations. To further explore this possibility, we searched for a direct link between binaries and BSSs. In agreement with previous studies[5,16,32,40] in globular and open clusters, our dataset shows a positive and strong correlation (with a Pearson coefficient 0.81) between the fraction of binaries and that of BSSs (Figure 4).

We used this tight relation to mathematically evaluate whether the correlations found for the BSS frequency (shown in Fig. 2) can be due to those observed for the binary fraction (Fig. 3). To this end, we re-wrote all the latter (see Supplementary Table 1) in terms of $\log(F4_{BSS})$, by substituting to $\log(f_{bin})$ the best-fit relation between these two quantities (Fig.4 and Supplementary Table 1). The results are impressive, and Figure 5 graphically shows those obtained for $\rho_0$, $\Gamma_{S+S}$ and $A^+$: the computed relations (grey solid lines) are almost perfectly superposed to the best-fit relations (black dashed lines). Notably, if the same procedure is applied to other sub-populations (such as HB and MSTO stars) in place of BSSs, the obtained results are totally inconsistent with the best-fit relations between the specific frequency of these stars and the cluster environmental parameters (see Supplementary Fig. 5 and Discussion).

The observed correlations, on their own, cannot be considered as irrefutable proofs of the physical connection between BSSs and binaries. However, the fact that the correlations observed for binary systems properly reproduce those found for BSSs (Figure 5), while they fail if other sub-populations are considered, supports the BSS-binary link. Thus, in the light of the current knowledge of the BSS phenomenon, the conclusion that BSSs are mainly originated from binary systems and follow the binary demography in different environments looks like the cleanest interpretation of the obtained results. This conclusion is further consolidated by the plots shown in Supplementary Fig. 6 that demonstrate how the detected trends are completely erased when the BSS specific frequency is divided by the binary fraction.

The emerging scenario suggests that the overall binary content is affected by the environmental conditions, with low-density environments being naturally effective in forming and/or preserving more primordial binaries (and binary by-products), compared to high-density environments. Indeed, a substantial decrease of the binary fraction with the environmental density was recently suggested to explain the marked decrease of fast rotating BSSs observed in high-density clusters[41] and additional empirical confirmations are coming from systematic multi-epoch spectroscopic studies of binary systems in star clusters. The exploration[42,43] of the central region of the GGCs NGC 3201 and 47 Tucanae has confirmed that the BSS population contains a fraction of binaries that is 3-5 times larger than the cluster average binary content, fully supporting a BSS binary origin. Moreover, the comparison between the ratio of binary to single BSSs ($R_{B/S}$) in the two systems demonstrates that the low-density cluster NGC 3201 has a ratio ($R_{B/S}$=1.35) that is 10 times larger than that of the high-density cluster 47 Tucanae ($R_{B/S}$ =0.12). In the present framework, this can be read as the natural consequence of the overall smaller binary fraction in high-density environments.

In turn, while dozens of BSSs for each sampled luminosity are observed in low-density conditions, the BSS specific frequency significantly decreases to just a few units in clusters with high values of $\rho_0$, $\Gamma_{S+S}$ and $A^+$ (Figs. 2d,e,f). Significant work remains to be done to assess the fraction of BSSs that are formed through (binary mediated) dynamical collisions with respect to those formed via binary secular evolution, and the impact of the binary fraction on the overall collision rate. However, a first estimate can be obtained by using simplified analytical definitions[44] of the single-binary ($\Gamma_{S+B}$) and binary-binary ($\Gamma_{B+B}$) collision rate (see Section "Collisional parameters involving binaries" in Methods). The resulting dependence of $F4_{BSS}$ on these parameters is shown in Figure 6. As apparent, the dependence of $F4_{BSS}$ on $\Gamma_{S+B}$ is shallower (and less significant) than the trend found with $\Gamma_{S+S}$ (see Fig. 2e), and the BSS specific frequency becomes almost independent of the collisional parameter when binary-binary interactions ($\Gamma_{B+B}$) are considered. This preliminary analysis, although simplified, shows that no evidence of positive trend between $F4_{BSS}$ and the collision probability emerges from the available data, even when binary-mediated encounters are considered. The same results are obtained also if considering the number of core BSSs (in place of $F4_{BSS}$), at odds with what found in numerical simulations[14] that predict a positive correlation. This clearly indicates that new theoretical investigations and simulations are now needed to properly describe the effects of the environment on binary systems and their role in generating BSSs.

# Discussion

During the first decade of the 2000's many investigations have been devoted to identifying correlations between the BSS content and the physical properties of GGCs. Although the adopted photometric catalogues, BSS selections (in terms of radial distance from the centre, magnitude limits, etc.), definitions of the BSS frequency (or BSS number), and cluster parameters never are exactly the same in the different works, the main findings can be summarized as follows.

BSSs vs. cluster mass/luminosity – The relation between the BSS specific frequency and the total luminosity (mass) of the parent clusters was detected[3,20-23] as soon as the first suitable BSS catalogues became available, and it is fully confirmed here (see Fig. 2a). A possible anticorrelation with the central velocity dispersion of the host cluster was also suggested[20,21], but the kinematic data available at that time were not accurate enough to clearly set it. Our study now shows an irrefutable anticorrelation between $F4_{BSS}$ and $\sigma_0$ (Fig. 2b). A strong sub-linear, correlation has been found[23,32,45] between the *number* of *core* BSSs ($N_{BSS,core}$) and the cluster core mass (derived from the core luminosity). Although the present work analyses the fraction of BSSs observed in the whole WFC3 field of view, the number of core BSSs in our sample confirms this finding (with a slightly steeper slope of 0.49, in place of 0.40).

BSSs vs. relaxation times, central density, and collisional parameter – Some non-significant hints of anti-correlation of the BSS specific frequency with the central and the half-mass relaxation times have been reported[20,21]. Here we find, instead, a positive, although not very significant trend, with $t_{rc}$, and a (clearer) anti-correlation with the $A^+$ parameter (see Supplementary Fig. 4). Controversial or poorly significant results have been previously found about the dependency of the BSS frequency on the cluster central density, with the detection[3] of some hints of anti-correlation, especially when limiting the sample to the densest clusters[22] or to the BSSs outside the core[21]. The present work finally shows that $F4_{BSS}$ strongly anti-correlates with the central surface brightness and density (Figs. 2c,d). Previous studies[20,21] also suggested a possible anti-correlation between the specific frequency of core BSSs and the collisional parameter, or at least a drop of the number of BSSs per unit luminosity in highly collisional systems[3,22,33], while here we show the existence of a strong and clear anti-correlation between $F4_{BSS}$ and $\Gamma_{S+S}$ (Fig. 2e). The number of core BSSs has been previously found[32] to positively correlate with the collision probability (see also [23] for the densest core clusters in their sample, and [14] for results from numerical simulations), while here we find that both the specific frequency (Supplementary Figure 1b) and the number of core BSSs anti-correlates with $\Gamma_{S+S}$.

BSSs vs. binaries – A direct and significant correlation between the core BSS frequency and the core binary fraction was found[40] for a sample of low-density clusters. A positive correlation between these two parameters has been subsequently detected[16] also for higher density systems, although the post-core collapse clusters in the sample clearly deviate from it. An increasing trend of $N_{BSS,core}$ with the number of core binary systems, although much less significant than the one with the core mass, has been also reported[32,45]. The relation shown in Fig. 4 now solidly confirms the existence of a positive correlation between $F4_{BSS}$ and $f_{bin}$ for clusters of any density, including the post core-collapsed ones (triangles), and considering the overall binary fraction. On the other hand, previous work has demonstrated that the other low-density Galactic environments where BSSs have been detected (the halo, disk, open clusters) are all characterized by very high binary frequencies[46-49], with 76% of the BSSs in the open cluster NGC 188 being in binary systems[4]. Moreover, recent spectroscopic investigations of the binary content in globular clusters have shown that the fraction of binary BSSs is higher in low-density (NGC 3201)[42] than in high-density (47 Tucanae)[43] conditions.

In summary the results presented in this work confirm some previous findings (as the BSS dependence on the sampled luminosity and the total cluster luminosity/mass), allow the clear definition of previously suspected dependences (as the one with the central velocity dispersion), and provide new correlations (as those with the cluster central surface brightness, central density, level of collisionality, and dynamical age). This has been possible thanks to the superior quality of the BSS sample discussed here in terms of homogeneity (see "Constructing the normalized CMD" above), and UV-sensitivity (see "Data analysis" and "On the completeness of the samples"), combined with an efficient cleaning procedure (see "Differential reddening correction and proper motion selection").

We emphasize that the relation shown in Fig.1b demonstrates the necessity of normalizing the number of BSSs to the sampled luminosity when searching for correlations with the cluster parameters. In fact, this relationship implies that the observed number of BSSs can differ from one cluster to another just because the observations have sampled different luminosities and, if not taken into account, the $N_{BSS}$-$L_{samp}$ dependence can "muddle" or even delete any other correlation. To convince the reader of the problem, let's consider the case of two high-density clusters, NGC 7078 and NGC 6752. They have large and similar central density, $\log(\rho_0)$ equal to about 5, but they are very different in terms of sampled luminosity, hence in BSS content (see Supplementary Data 1): $L_{samp}$ amounts to $30.6\times10^4$ $L_\odot$ and $8\times10^4$ $L_\odot$ in NGC 7078 and NGC 6752, respectively, and the former hosts 111 BSSs, while the latter has only 22. These values (found in clusters with the same value of $\rho_0$) are at the extreme high- and low- ends of the observed distribution of $N_{BSS}$. Hence, they would contribute to disarrange any possible correlation between $N_{BSS}$ and $\rho_0$. On the other hand, the BSS specific frequency is similar in these two clusters (3.6 and 2.7, respectively), indicating a low production/survival efficiency of BSSs in similarly dense environments, and well collocating both systems in the right-bottom corner of Fig.2d, where they contribute to delineate the observed $\log(F4_{BSS})$- $\log(\rho_0)$ relation. Thus, to reliably explore the BSS dependency on any other cluster properties, we first need to "remove" the trend between $N_{BSS}$ and the sampled luminosity, and this is what $F4_{BSS}$ does.

Instead of using the BSS specific frequency, different works[21,23,32,33,45] in the literature have searched for correlations between the number of BSSs in a given cluster region (such as the core) and the host parameters. However, the number of core BSSs in the example mentioned above still remains quite different (38 in NGC 7078 and just 8 in NGC 6752), in spite of the same cluster central density. Moreover, the definition of cluster core is by itself hazardous in high-density clusters (and even meaningless in post-core collapsed systems) and it poses strong limitations in the derived sample size. For instance, 35% of the clusters investigated in [23] host less than 10 BSSs and approximately 10% have less than 4 BSSs (with even zero BSSs in the post core collapse system NGC 6284). Not only these are statistically poor BSS samples, but they are also found in the clusters with the lowest core mass and with the highest density. Hence, as matter of fact, they dominate the investigated correlations and strongly contribute to determine the slope of the trend with the core mass.

While the normalization to a quantity that accounts for the sampled luminosity (by using $L_{samp}$ or the number of HB or MSTO stars; see Supplementary Figure 3) is a necessity when searching for any existing secondary correlations, it clearly introduces a sort of additional dependence because all the cluster properties and environmental parameters that can be derived (including the core mass or the number of binaries) do depend on luminosity. Thus, we are in the situation where all the sub-populations, by them-selves, have an intrinsic dependence on luminosity (a dependence that can differ from a population to another), and the normalization to $L_{samp}$ introduces an additional dependence. However, the fact that the correlations with the environmental parameters obtained for the specific frequency of MSTO and HB stars are highly different from those observed for BSSs and binaries testifies that this secondary dependence is not dominant.

In this paper, the evidence that the trends between $F4_{BSS}$ and the environmental parameters (Fig. 2) are reproduced by those observed for the binary fraction (Fig.3) combined with the $f_{bin}$-$F4_{BSS}$ relation (Fig. 4) is interpreted as support to the link between binaries and BSSs. In the following, we address instead the possibility that binaries and BSSs are actually unrelated, and that the derived trends are due to the underlying dependence on luminosity introduced by the normalization of $N_{BSS}$ to $L_{samp}$. If so, we should expect that analogous results are obtained if any other binary-unrelated population (such as HB or MSTO stars) is adopted instead of BSSs. To test this possibility, we computed the MSTO and the HB specific frequencies ($F4_{MSTO}$ and $F4_{HB}$) as the number of MSTO and HB stars normalized to $L_{samp}$ in units of $10^4$ $L_\odot$, respectively. Then we combined the trends observed for the binary fraction (Fig. 3) with the $f_{bin}$-$F4_{MSTO}$ relation, and thus determined the expected dependence between $F4_{MSTO}$ and the environmental parameters. The result obtained for the dependence on the central density $\rho_0$ is shown in Supplementary Figure 5a. Clearly, the derived relation (red line) is totally inconsistent with the best-fit one (black line), demonstrating that it is not possible to reproduce the trend between $F4_{MSTO}$ and $\rho_0$ by combining the $f_{bin}$ -$\rho_0$ relation with the $f_{bin}$-$F4_{MSTO}$ one. The same holds if HB stars are considered (see Supplementary Figure 5b), and with respect to any other environmental parameter. Thus, as matter of fact, the relations found between binaries and the environment are able to reproduce only the BSS-environment relations. This evidence naturally supports the conclusion that the relations linking the BSS frequency to the cluster properties are driven by the demography of binary systems in different environments.

# Methods

**Data analysis**

The photometric analysis of the HST images used in this study (see Supplementary Information) has been specifically optimized for the BSS search: it followed a UV-guided photometric approach[2,25,27,50], where stellar objects are identified in the F275W images. Indeed, in old stellar systems like GGCs, the optical emission is dominated by cool red giant branch (RGB) stars, while these become particularly faint, and BSSs are among the brightest objects at UV wavelengths. Hence, the usual problems associated with photometric blends and crowding in the high-density central regions of GGCs are minimized, and BSSs can be reliably recognized and easily separated from both other evolved and unevolved populations at UV wavelengths. This allows the selection of complete BSS samples (see "On the completeness of the samples" below), from which reliable star counts can be obtained. To measure the magnitudes of each detected source, a Point Spread Function fitting (PSF) procedure was performed, by using a set of PSF libraries specifically perturbed to take into account both the spatial and the temporal PSF variations. The stellar positions have been corrected for geometric distortion and transformed into the absolute astrometric system using the stars in common with the most updated Gaia Data Release[51].

**On the completeness of the samples**

The BSS sample discussed here represents the most complete collection of BSSs ever observed in GGCs. It is based entirely on UV observations, which maximize the detection of blue/hot populations in high-density environments. In previous works we have discussed the clear advantage of selecting BSSs in the UV-CMD, constructed by using the UV guided approach. In [25] we have discussed the illustrative cases of a few high-density clusters, comparing the BSS selections in the UV band, with those performed in the optical and previously published for these same systems, clearly demonstrating the great advantage of a UV-guided search for BSSs. As discussed in [25] artificial star experiments have been performed in a few (intermediate and high-density) cases always obtaining completeness levels larger than 80%. We also verified that the ratio between the number of MSTO stars (the faintest population considered in this study; see the selection box in Figures 2, 3, 4 in [25]) and the number of HB stars (the brightest ones) stays constant independently of the cluster central density (see Supplementary Figure 7). This further confirms that even a sub-population fainter than the BSS samples considered in this study is affected by a negligible level of incompleteness.

**Differential reddening correction and proper motion selection**

To optimize the definition of the evolutionary sequences in the CMD and allow a safe selection of the BSS population in each cluster, the catalogues obtained from the photometric analysis were corrected for differential reddening. In fact, the dust extinction can be highly variable, even on scales of just a few arcseconds, especially in the direction of GGCs affected by large values of reddening. This can severely distort the evolutionary sequences in the CMD, preventing the proper identification of the stellar populations hosted in the system. To correct for this effect, we computed the amount of differential reddening following the procedure described in previous papers[52], where the reddening of each star is determined by measuring the displacement of its local CMD (i.e., the CMD built from the spatially closest objects) with respect to the cluster mean ridge line.

Although Galactic field contamination is expected to be low in the investigated clusters, because they are mainly located in the Milky Way halo and the analysis is limited to their central regions (r<85”) where the cluster member population is always largely dominant, a proper motion analysis has been used to identify and exclude possible field interlopers from the BSS samples. To this end, we cross-correlated the UV-photometric catalogues with the catalogues of relative proper motions recently published[28] for the GGCs observed in the HST UV Legacy Survey of GGCs[24]. Typically, proper motions have been determined over a time-baseline of 7-8 yr using the data from the ACS Survey of GGCs (GO-10775) as first-epoch observations. To separate cluster members from field stars, we built the vector point diagrams (VPDs) plotting all the available measures in each cluster. By construction, cluster members define a well-grouped distribution clumped around (0,0) in the VPD, while field contaminants generally exhibit a much more scattered distribution. Thus, we assumed as member BSSs all those included within 5 times the dispersion of the proper motion distribution drawn by the stars in the same magnitude range. We conservatively kept into the sample also the BSSs with no proper motion measure. After this field decontamination procedure, the total BSS sample is reduced from 3747 to 3419 objects, implying a global effect of the order of 10%. We also verified that the fraction of BSSs with no PM measure is below 15% in the large majority of the clusters, and the results remain unchanged even when removing the systems with the poorest PM analysis.

**Constructing the normalized CMD**

To perform a homogeneous selection of BSSs in clusters with different metallicities, reddening and distances, we followed the same approach already tested in previous studies[25,27,39], which consists in constructing the "normalized CMD (n-CMD)" and using the same selection box in all systems. The 48 clusters considered in this study cover the entire range of metallicities spanned by the GGC system, from [Fe/H]= −2.4 to [Fe/H]= −0.4. Hence, they have been divided in 7 groups according to their metallicity. For each group, a 12 Gyr-old isochrone with the corresponding value of [Fe/H] was adopted from the BASTI database[53] and arbitrarily shifted until the MSTO point was located at magnitude and colour equal to zero. Then, the observed CMD of every cluster has been shifted to match the reference isochrone of the proper metallicity group. In this way, the BSSs belonging to each cluster result to populate the same region of the n-CMD, still satisfying the original definition of "stragglers" as stars with brighter magnitudes and bluer colours than the parent cluster MSTO, and a unique colour-magnitude box is used to select the population.

**Sampled luminosity and sampled populations**

We estimated the light sampled by the observations in each cluster by integrating, between 0" and 85" from the centre, the King model that best reproduces the projected star density profile of the cluster. In doing this we adopted the cluster structural parameters listed in Supplementary Data 1. Most of them are taken from the Harris compilation[19] or from [54]. The integration was performed by imposing that the integral of the King model between 0 and the cluster tidal radius equals the total cluster luminosity derived from the observed integrated V magnitude. The sampled luminosities in units of $10^4$ $L_\odot$ are listed in Supplementary Data 1. It is important to emphasise that, although the adoption of the sampled luminosity optimizes the trends plotted in Figure 2, analogous relations are found also by adopting different normalizations of the BSS specific frequency. This is shown both in Supplementary Figure 2, where the number of BSSs is normalized to the sum of the luminosities of all the stars detected in the field of view, and in Supplementary Figure 3, where the adopted normalization is the number of "normal" cluster stars, either the HB population, or the MSTO and sub-giant branch stars (the latter have been selected according to the box shown in Figures 2, 3, and 4 of [25], comprising the brightest portion of the MSTO and a first segment of the sub-giant branch).

**The single-single collisional parameter**

A standard way to estimate the level of collisionality of a cluster is through the collisional parameter[31] ($\Gamma_{S+S}$), which provides an estimate of the collision probability between single stars and here is intended as a generic indicator of collisionality. The values adopted in this study have been determined as $\Gamma_{S+S} \propto \rho_0^{1.5}\, r_c^2$, where $\rho_0$ is the central luminosity density in units of $L_\odot$ pc$^{-3}$, $r_c$ is the core radius in parsecs, and the constant of proportionality has been neglected (set equal to unity) because we are just interested in the relative ranking of the target systems. The central density has been estimated by following equation (7) in [55], and the necessary cluster parameters (such as the distance, reddening, integrated magnitude, King concentration parameter and core radius) have been adopted from previous studies[19,54]. The anti-correlation between the BSS specific frequency and $\Gamma_{S+S}$ remains almost unchanged even if other definitions[16,56] of the collisional parameter or central density are used. For instance, we considered the specific encounter frequency ($\gamma$), that according to its definition (the collisional parameter normalized to the cluster mass[56]) represents a measure of the chance that a particular star in a globular cluster undergoes an encounter, and we verified that the discovered relations remain valid. We also searched for secondary effects that could be hidden behind the main dependence. In particular, we computed the residuals $\Delta\log(F4_{BSS})$ with respect to the best-fit relation in the $\log(f_{bin})$-$\log(F4_{BSS})$ diagram and investigated their trend with the collisional parameter. The result is shown in Supplementary Fig. 8: no positive trend with $\Gamma_{S+S}$ is found; conversely, we find a weak (if any) but negative trend, providing possible evidence that, for a given binary fraction, a higher collision rate might suppress BSS formation. Further considerations, about the role of single-binary and binary-binary collisions[14,44], are discussed in Section "Collisional parameters involving binaries" below.

From the observational point of view, the most likely sub-population of collisional BSSs has been identified so far only in a few post-core collapse clusters showing two separated and well-defined sequences in the CMD[57-61], the bluest one being consistent with the location predicted for collisional by-products[11]. This could suggest that collisions are particularly effective in producing BSSs mainly during the central density increase that occurs at the epoch of core collapse and the gravothermal oscillation phase characterizing the latest stages of star cluster dynamical evolution. However, the observed trends show no evidence of this additional component: the anti-correlations shown in Figs. 2d,e,f remains essentially unchanged if the post-core collapsed systems are excluded from the sample, thus indicating that the collisional component (if any) is certainly not abundant enough to compensate the decrease of the overall BSS specific frequency due to the deficit of binaries.

**Measuring dynamical evolution with the BSS dynamical clock**

BSSs are thought to be significantly more massive than the other shining stars in GGCs. For this reason, they are expected to migrate toward the cluster centre faster than their lighter sisters, under the effect of dynamical friction. This is the core concept of the so-called "dynamical clock"[25,36], an empirical tool that uses the observational properties of BSSs to measure the level of internal dynamical evolution of star clusters[25,38,39]. In particular, the level of BSS central segregation with respect to normal (lighter) stars has been quantified via the $A^+$ parameter[34,35], defined as the area between the cumulative radial distribution of BSSs and that of a lighter population, such as, e.g., RGB or HB stars. Thus, this parameter turns out to be small ($A^+$ approximately 0-0.05) for dynamically-young stellar systems, measuring a negligible difference between the two radial distributions due to the fact that dynamical friction has not yet significantly segregated BSSs into the cluster centre, while it reaches values larger than 0.30 for post-core collapsed systems. The $A^+$ parameter has been found to correlate with the number of the present-day central relaxation times experienced by the host cluster ($t_{rc}$, the parameter commonly used to quantify the dynamical ageing), thus demonstrating that both these quantities measure the cluster dynamical evolution. However, while the computation of $t_{rc}$ suffers from a number of simplifying assumptions and approximations, the $A^+$ parameter provides a direct and fully observational measurement of the degree of mass segregation developed during the entire cluster's evolution. As shown in Supplementary Figure 4, the anti-correlation between $F4_{BSS}$ and the parent cluster dynamical ageing turns out to be much sharper when the $A^+$ parameter is used instead of $t_{rc}$, once more demonstrating the advantage of using the BSS dynamical clock to measure the dynamical age of stellar systems.

**The binary fraction**

The binary fractions adopted in this work are the values of $f_{bin}^{q>0.5}$ listed in Table 2 of [16] for the ACS Wide Field Camera (WFC) field. For each cluster, they have been determined as the ratio between the number of stars observed in an offset region of the CMD parallel to the MS, where the photometric blends of binary components with mass ratio q>0.5 are expected to lie, normalized to the total number of MS stars in the same magnitude range. The spatial region is the same as the one where $F4_{BSS}$ has been determined, namely, the HST ACS/WFC field of view. We verified that using the binary fractions estimated for mass ratios $q>0.6$ and $q>0.7$ (which are also provided in [16]) has a negligible impact on the discussed correlations. These binary fractions, which are photometrically estimated along the MS, do not correspond to the total binary frequencies in star clusters, nor they bring information about physical parameters (e.g., the orbital separation) that can impact the probability of mass transfer and collisions. In addition, they have been estimated[16] from stars characterized by completeness levels above 50%. Nevertheless, the evidence that the correlations between $f_{bin}$ and the cluster parameters so precisely explain the corresponding trends observed for $F4_{BSS}$ (see Fig. 5) strongly suggests that the missing portion of the binary fraction plays a negligible role in the BSS characterization. A few relations between the binary fraction and cluster parameters are discussed in [16]. Overall, the core binary fraction shows a strong correlation with $M_V$, less significant anti-correlation with $\sigma_0$ and $\rho_0$, a mild anti-correlation with the collisional parameter, and non-significant anti-correlations with the central and the half-mass relaxation times.

**Collisional parameters involving binaries**

Mainly due to their larger cross section, binary systems have larger probabilities of gravitational encounters compared to single stars[12,14,15]. While dedicated theoretical investigations are needed to properly follow stellar collisions in different environments, we used simplified analytical expressions[44] of the single-binary and binary-binary collision rates ($\Gamma_{S+B}$ and $\Gamma_{B+B}$) to investigate how $F4_{BSS}$ scales with these parameters. According to the definition in [44], we assumed $\Gamma_{S+B} = 1/\tau_{S+B}$, where $\tau_{S+B}$ is the time between single–binary encounters as defined in their equation A10. Similarly, $\Gamma_{B+B} = 1/\tau_{B+B}$, where $\tau_{B+B}$ is the time between binary–binary encounters from their equation A8. In the computation of $\tau_{S+B}$ and $\tau_{B+B}$, we assumed 2 AU as average binary semi-major axis[32], 0.3 $M_\odot$ as average stellar mass in all clusters, and 0.3 $R_\odot$ as the corresponding stellar radius, and we neglected the possible presence of triple systems.

**Data availability**

The source data underlying all graphs presented within the main text and Supplementary Figures are provided as a Source Data File. Source data are provided with this paper. The data analyzed in this study are provided as photometric catalogues of each investigated cluster and can freely downloaded at the repository web page: http://www.cosmic-lab.eu/Cosmic-Lab/nCMDs_for_BSSs.html.

**Acknowledgements**

This research is part of the project COSMIC-LAB at the Physics and Astronomy Department of the Bologna University (see the web page: http://www.cosmic-lab.eu/Cosmic-Lab/Home.html ).

**Author contributions**

FRF designed the study and coordinated the activity. EV and BL provided expertise in dynamical evolution of star clusters. GP coordinated the acquisition of the database and DN, ML, CP, MC, ED and GB performed the data analysis. FRF and BL wrote the first draft of the paper. All the authors contributed to the discussion of the results and commented on the manuscript.

**Competing Interests**

The authors declare no competing interests.

**Correspondence to**

Correspondence and requests for materials should be addressed to F. R. Ferraro **francesco.ferraro3@unibo.it**

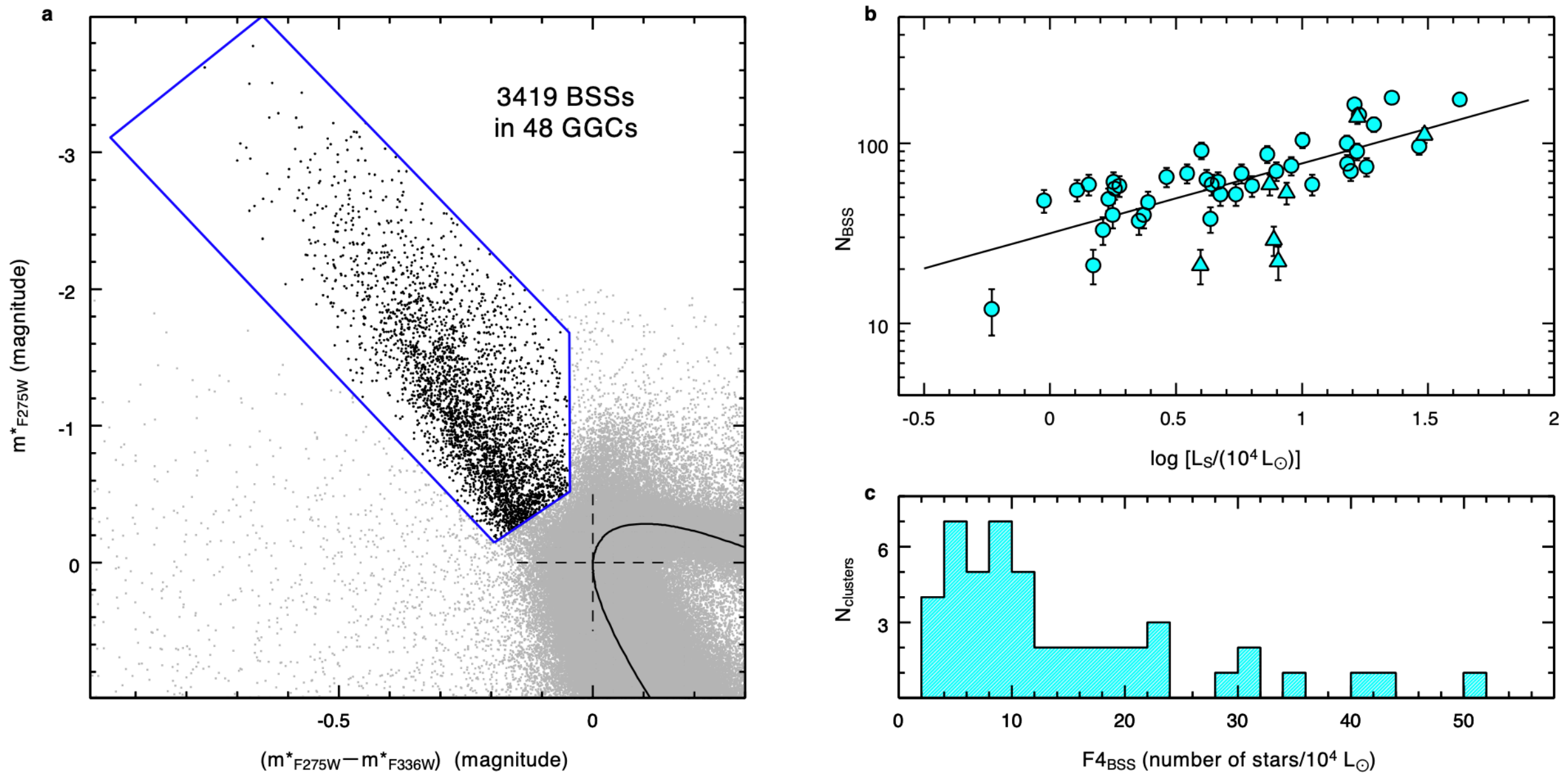

**Fig. 1 | The total sample of blue straggler stars (BSSs). a:** The entire sample of 3419 BSSs identified in 48 Galactic globular clusters are shown (black dots) in the "co-added" normalized colour-magnitude diagram (n-CMD, grey points) in the ultraviolet filter F275W and F336W. The BSS selection box is shown in blue. A 12 Gyr old isochrone at intermediate metallicity ([Fe/H]= −1.6) from the BaSTI library[53] is also shown (black solid line) for reference. The large dashed cross marks the location of the main sequence turnoff (at magnitude and colour equal to 0) in the n-CMD. This is the point at which all the observed CMD have been shifted. **b:** The number of BSSs ($N_{BSS}$) counted in each cluster increases as a function of the sampled luminosity (in units of $10^4\ L_\odot$). Post-core collapsed clusters (from [19] compilation) are plotted as triangles (here, and in all subsequent figures). The plotted errors (1 s.e.m) have been computed following the Poisson statistics (in most of the cases they are smaller than the symbol size). **c:** The distribution of the specific frequency $F4_{BSS}$ (defined as the number of BSSs per unit of $10^4\ L_\odot$ sampled luminosity) varies by a factor of 20 in the 48 Galactic globular clusters under investigation. Source data are provided as a Source Data file.

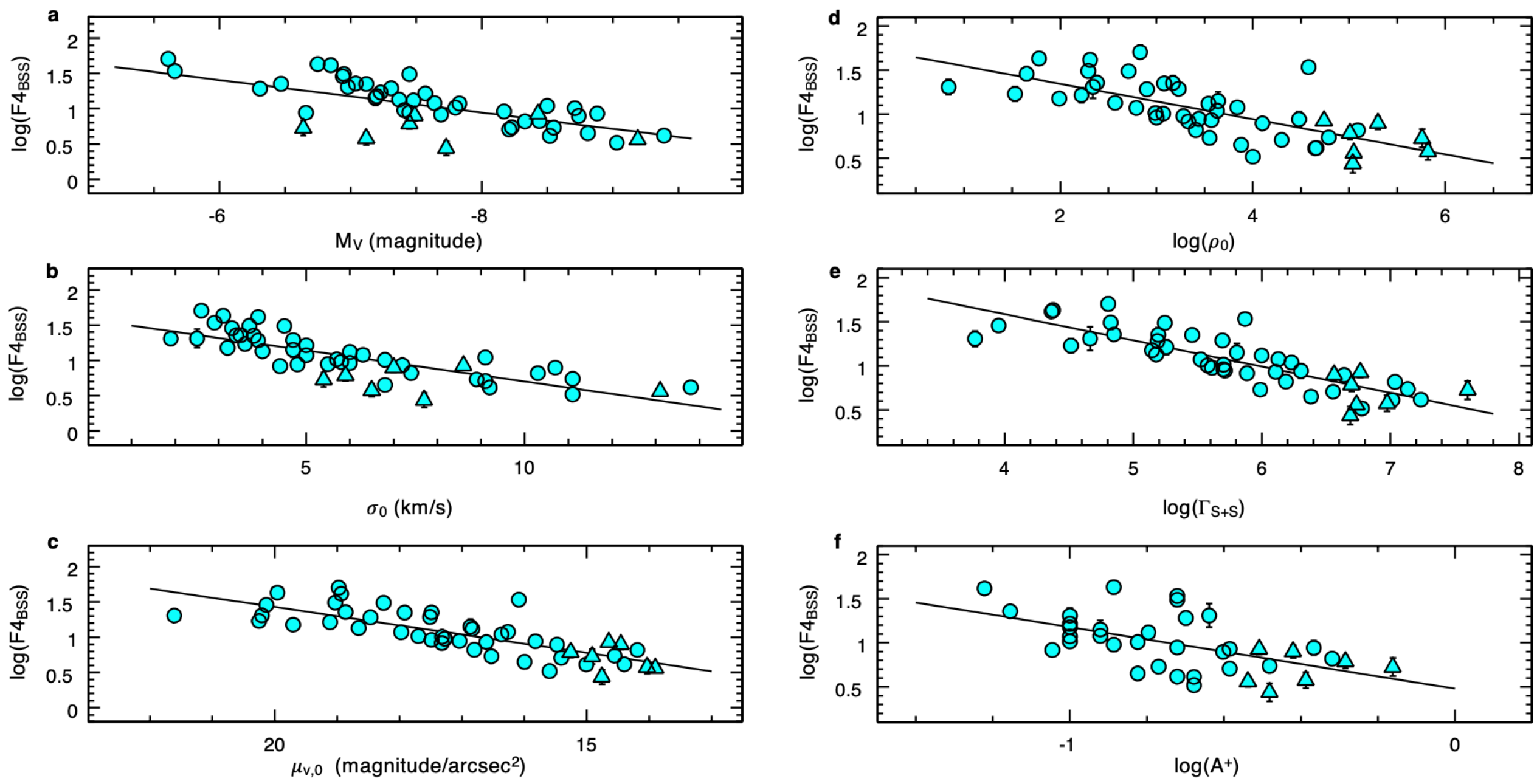

**Fig. 2 | The main relations linking the blue straggler stars (BSS) specific frequency ($F4_{BSS}$) to the cluster parameters. a:** The well-known relation between $F4_{BSS}$ (in number of stars/$10^4$ $L_\odot$) and the cluster total magnitude ($M_V$). **b, c, d, e, f:** The new relations emerging from this study, showing that $F4_{BSS}$ strongly anti-correlates also with the central velocity dispersion ($\sigma_0$), the central surface brightness ($\mu_{V,0}$), the central density ($\rho_0$, in units of $L_\odot/pc^3$), the single-single collisional parameter ($\Gamma_{S+S}$, in arbitrary units), and the dynamical age (parametrized by the $A^+$ parameter, in arbitrary units). In all the panels the best-fit relations to the data are shown as solid black lines. Their equations are provided in Supplementary Table 1. Errors are 1 s.e.m. Source data are provided as a Source Data file.

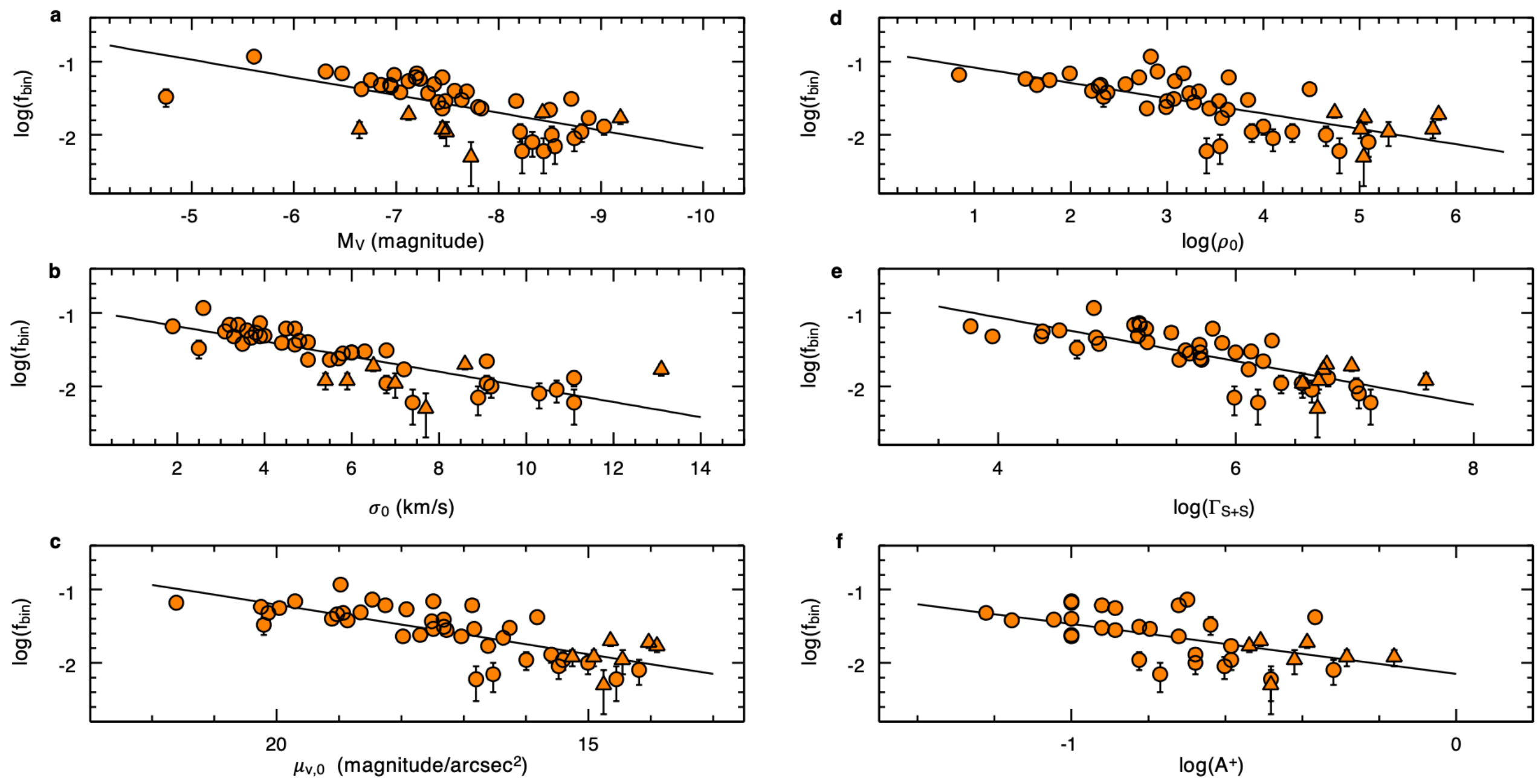

**Fig. 3 | The main relations linking the binary fraction to the cluster parameters.** As in Figure 2, but for the binary fraction ($f_{bin}$), instead of the blue straggler specific frequency: the binary fraction is plotted as a function of the cluster total magnitude ($M_V$), the central velocity dispersion ($\sigma_0$), the central surface brightness ($\mu_{V,0}$), the central density ($\rho_0$, in units of $L_\odot/pc^3$), the single-single collisional parameter ($\Gamma_{S+S}$, in arbitrary units), and the dynamical age (parametrized by the $A^+$ parameter, in arbitrary units), in panels a, b, c, d, e, f, respectively. Errors (1 s.e.m) are calculated by following the standard error propagation law, assuming Poissonian statistics for number counts (in most of the cases they are smaller than the symbol size). Source data are provided as a Source Data file.

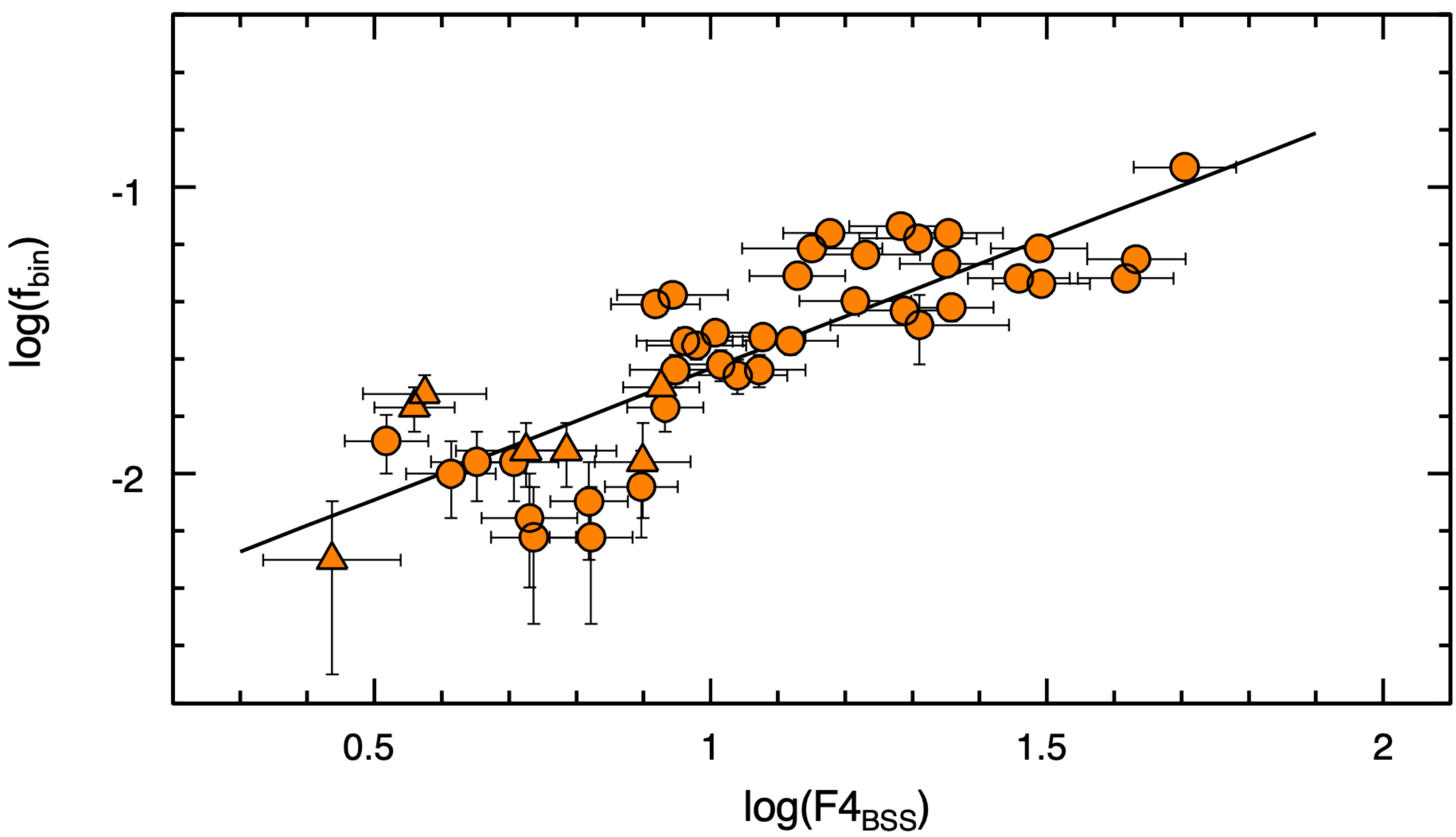

**Fig. 4 | The strong link between blue straggler stars and binaries.** The well-defined relation found between the fraction of binaries ($f_{bin}$) hosted by the parent cluster and the blue straggler specific frequency ($F4_{BSS}$, in number of stars/$10^4$ $L_\odot$) The best-fit relation to the data is shown as a solid line. Errors (1 s.e.m) are calculated by following the standard error propagation law. Source data are provided as a Source Data file.

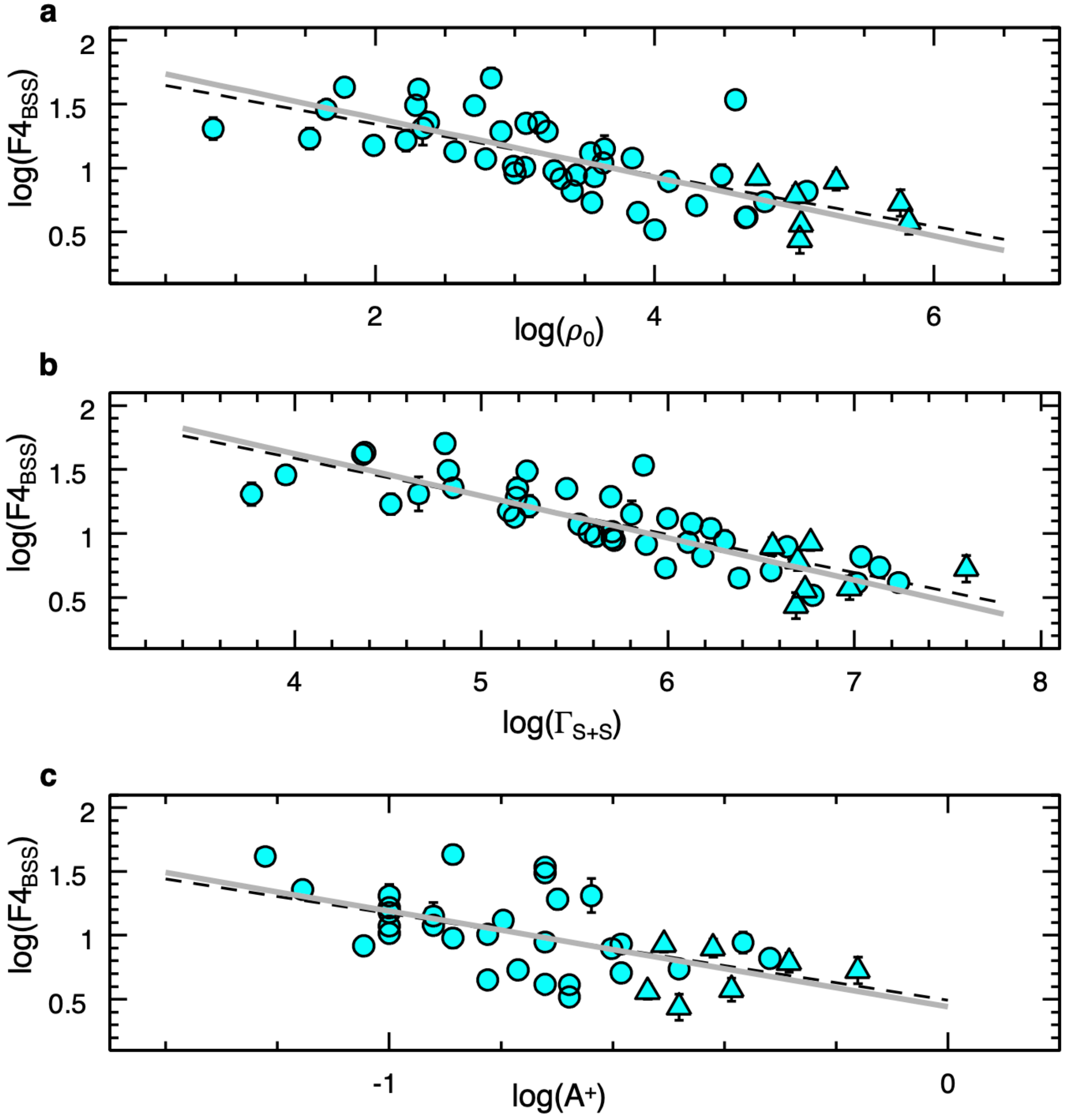

**Fig. 5 | The relations found between the blue straggler specific frequency and the cluster environmental parameters are driven by the binaries.** The figure shows that the trends plotted in Figure 2d,e,f between the blue straggler specific frequency ($F4_{BSS}$) and the cluster central density ($\rho_0$), collisional parameter ($\Gamma_{S+S}$), and $A^+$ parameter (black dashed lines in panels a, b, c, respectively) are very well reproduced (heavy grey lines) by combining the relation linking $f_{bin}$ and $F4_{BSS}$ (solid line in Figure 4) with the relations shown in Figures 3d,e,f for the binary fraction. $F4_{BSS}$ is in number of stars/$10^4$ $L_\odot$, $\rho_0$ in units of $L_\odot/pc^3$, the collisional and $A^+$ parameters are in arbitrary units. Errors are 1 s.e.m. Source data are provided as a Source Data file.

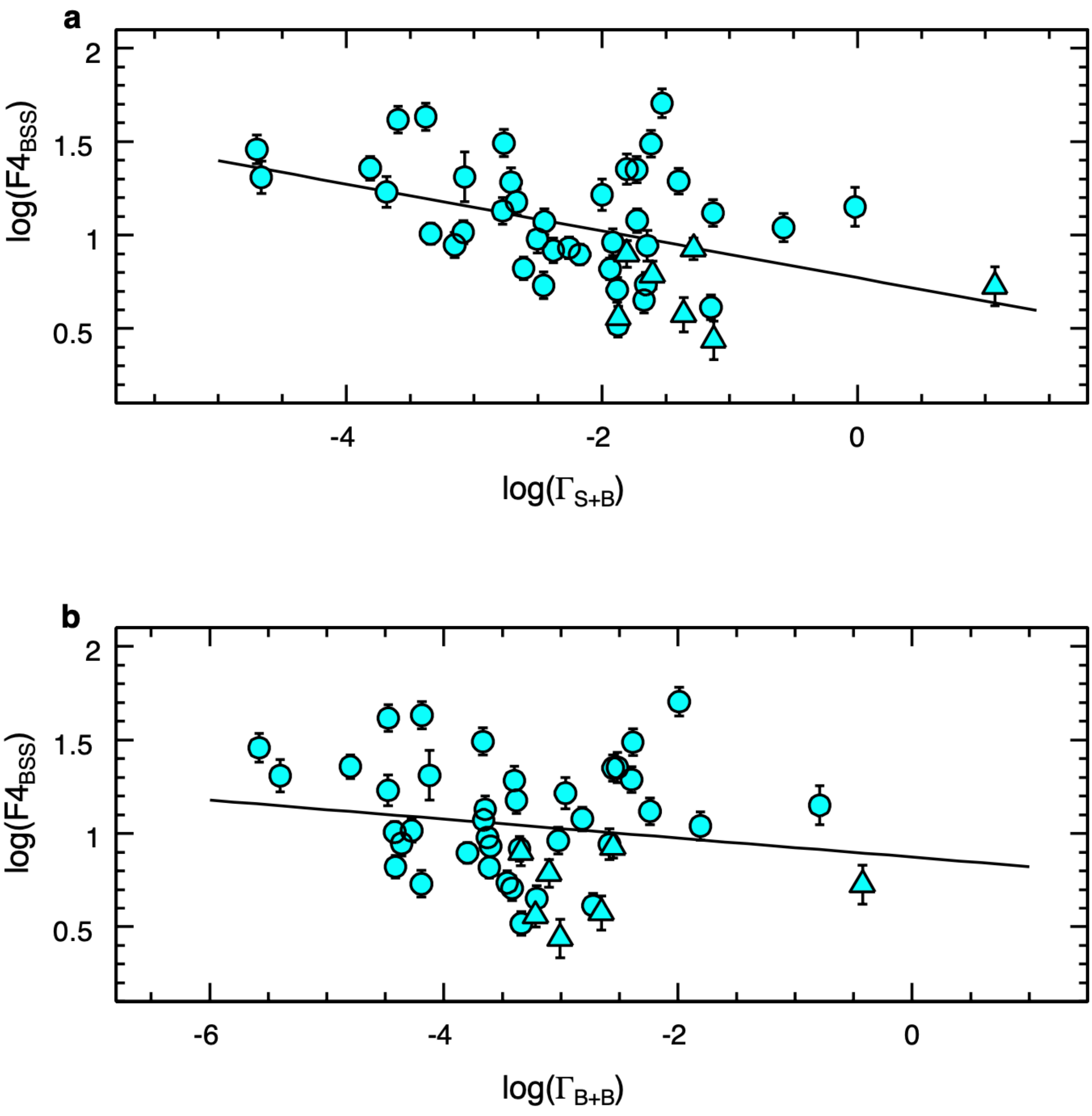

**Fig. 6 | The dependence of the blue straggler star specific frequency on the collisional parameters involving binaries.** The figure shows the blue straggler specific frequency ($F4_{BSS}$, in units of $L_{\odot}/pc^3$) as function of the probabilities[44] of dynamical encounters between binaries and single stars ($\Gamma_{S+B}$ in arbitrary units, panel a) and between binary systems ($\Gamma_{B+B}$ in arbitrary units, panel b). No evidence of positive trends emerges from the data. Source data are provided as a Source Data file.

# Supplementary Information for
# "A binary-related origin mediated by environmental conditions for blue straggler stars"

**The data set:** The present study is based on high-resolution images collected with the UVIS channel of the HST/WFC3 in the filters F275W (with pivot wavelength $\lambda_p$=2709.7 Å) and F336W ($\lambda_p$=3354.5 Å), acquired under the HST UV Legacy Survey of Galactic Globular Clusters[24] (GO-13297). The entire data set and data reduction procedures are described in detail in previous papers[25-27]. Here we briefly summarize the main points. The WFC3/UVIS consists of two chips, each of 4096 ×2051 pixels, with a pixel scale of 0.04 arcsec, resulting in a total field of view of about 162" × 162". The pointing was roughly centred on the cluster centre, thus allowing the uniform sampling of the innermost 85" of each observed cluster. In every band, different pointings were dithered by several pixels, and in some cases, they were also rotated by approximately 90° to allow an optimal subtraction of CCD defects, artifacts, charge loss and false detections. All the images have been corrected for the effect of poor charge transfer efficiency.

**The sample of clusters:** To minimize any non-homogeneity in the blue straggler star (BSS) selection, in the present analysis we used exclusively clusters observed with the same combination of UV filters and we excluded only a few systems for which the BSS population is not clearly distinguishable from main sequence turnoff stars (due to an insufficient signal-to-noise ratio at that magnitude level). We also required that the observations sampled a relevant fraction (> 80%) of the cluster core radius. Thus, we finally analysed a total of 48 Galactic Globular Clusters, corresponding to about 1/3 of the total Milky Way population and optimally sampling its entire parameter space, with central densities varying by several orders of magnitude (from log $\rho_0$ =0.5 to 6 in units of $L_\odot/pc^3$), absolute magnitudes ($M_V$) varying from −4.7 to −9, core radii ranging from approximately 0.03 to about 6 pc, and metallicities spanning the range [Fe/H] = −2.4, [Fe/H] = −0.4. The main parameters of the considered clusters are listed in Supplementary Data 1.

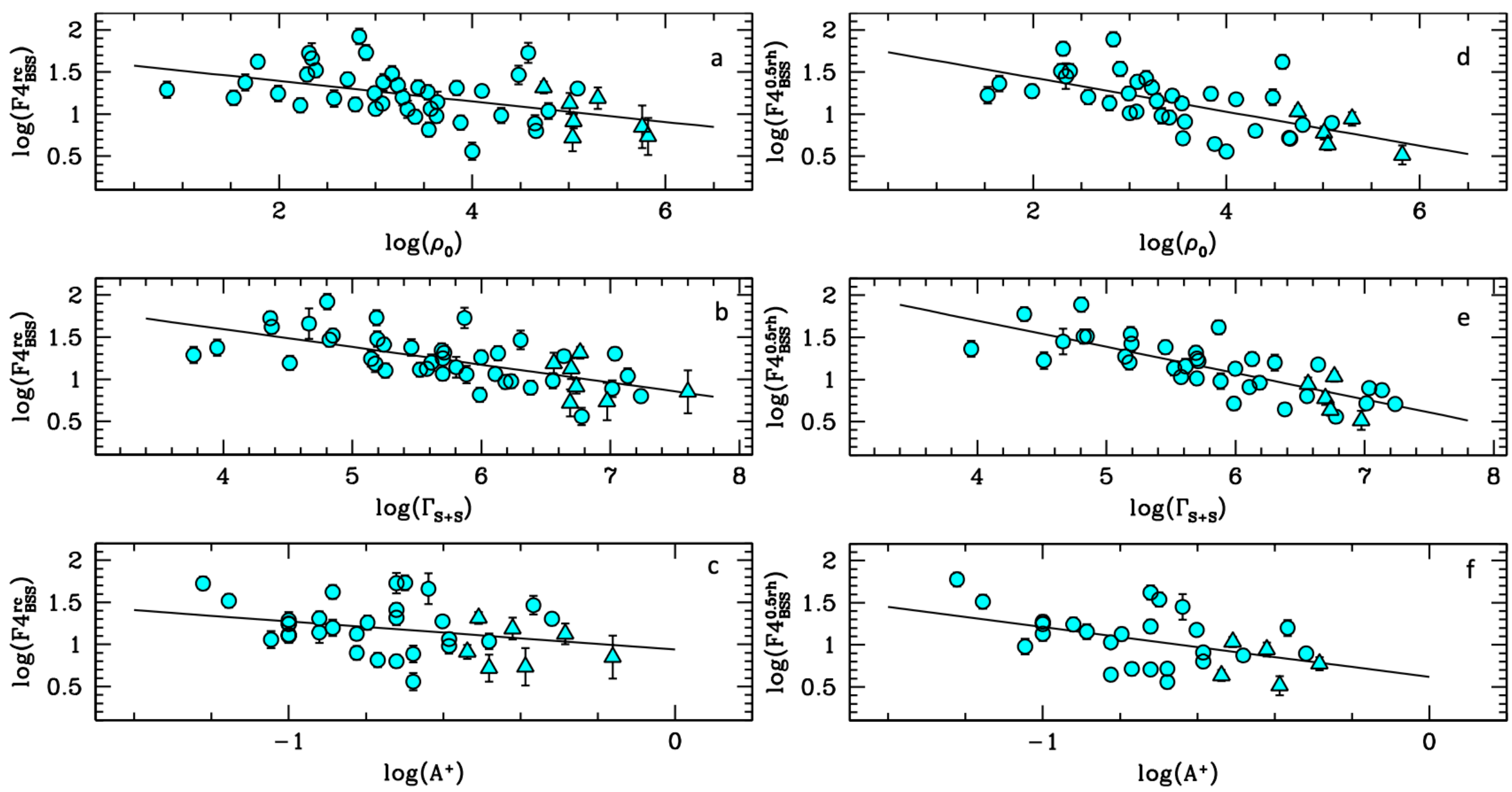

**Supplementary Fig. 1 | Changing the cluster region where the blue straggler frequency is computed.** The main relations shown in Figure 2d, e, f, with respect to the cluster central density ($\rho_0$ in units of $L_{\odot}/pc^3$, panels a, d), collisional parameter ($\Gamma_{S+S}$ in arbitrary units, panels b, e), and dynamical age parametrized by the $A^+$ parameter (in arbitrary units, panels c, f) persist even when the blue straggler specific frequency is computed within the cluster core ($F4^{rc}_{BSS}$, in number of stars/$10^4$ $L_{\odot}$, left panels) and one half the half-mass radius ($F4^{0.5rh}_{BSS}$, in number of stars/$10^4$ $L_{\odot}$, right panels). In all the panels the best-fit relations to the data are shown as solid black lines. Post-core collapse clusters (from [19] compilation) are plotted as triangles (here and in all subsequent figures). Errors are 1 s.e.m. Source data are provided as a Source Data file.

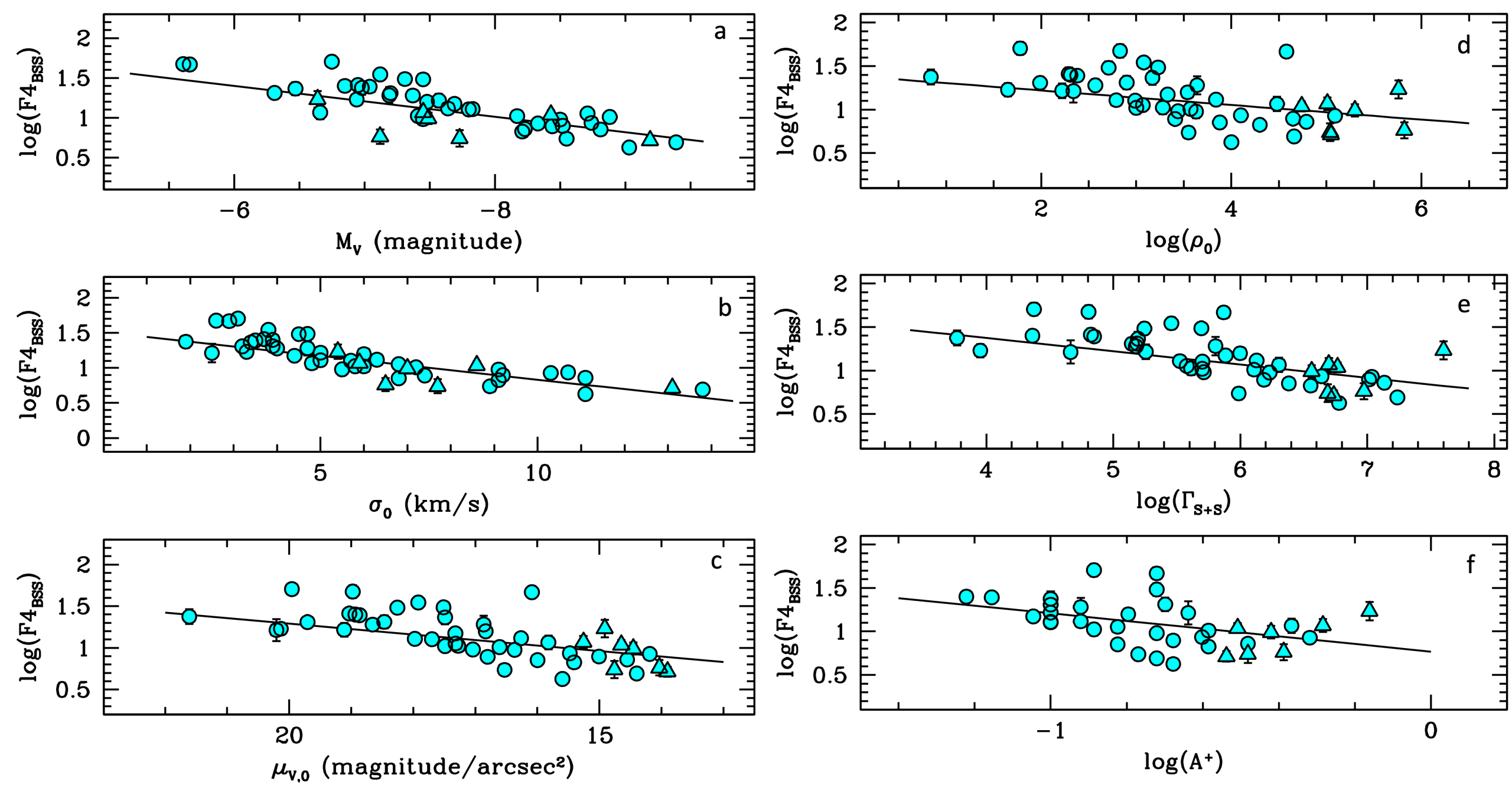

**Supplementary Fig. 2 | Changing the way of determining the sampled luminosity.** The relations shown in Figure 2 persist even when the blue straggler specific frequency ($F4_{BSS}$, in number of stars/$10^4$ $L_\odot$) is normalized to the luminosity computed by summing up the luminosities of all the stars detected in the field of view. Panels a, b, c, d, e, f show, respectively, the newly computed $F4_{BSS}$ as a function of the cluster total magnitude ($M_V$), the central velocity dispersion ($\sigma_0$), the central surface brightness ($\mu_{V,0}$), the central density ($\rho_0$, in units of $L_\odot$/pc$^3$), the single-single collisional parameter ($\Gamma_{S+S}$, in arbitrary units), and the dynamical age (parametrized by the $A^+$ parameter, in arbitrary units). In all the panels the best-fit relations to the data are shown as solid black lines. Errors are 1 s.e.m. Source data are provided as a Source Data file.

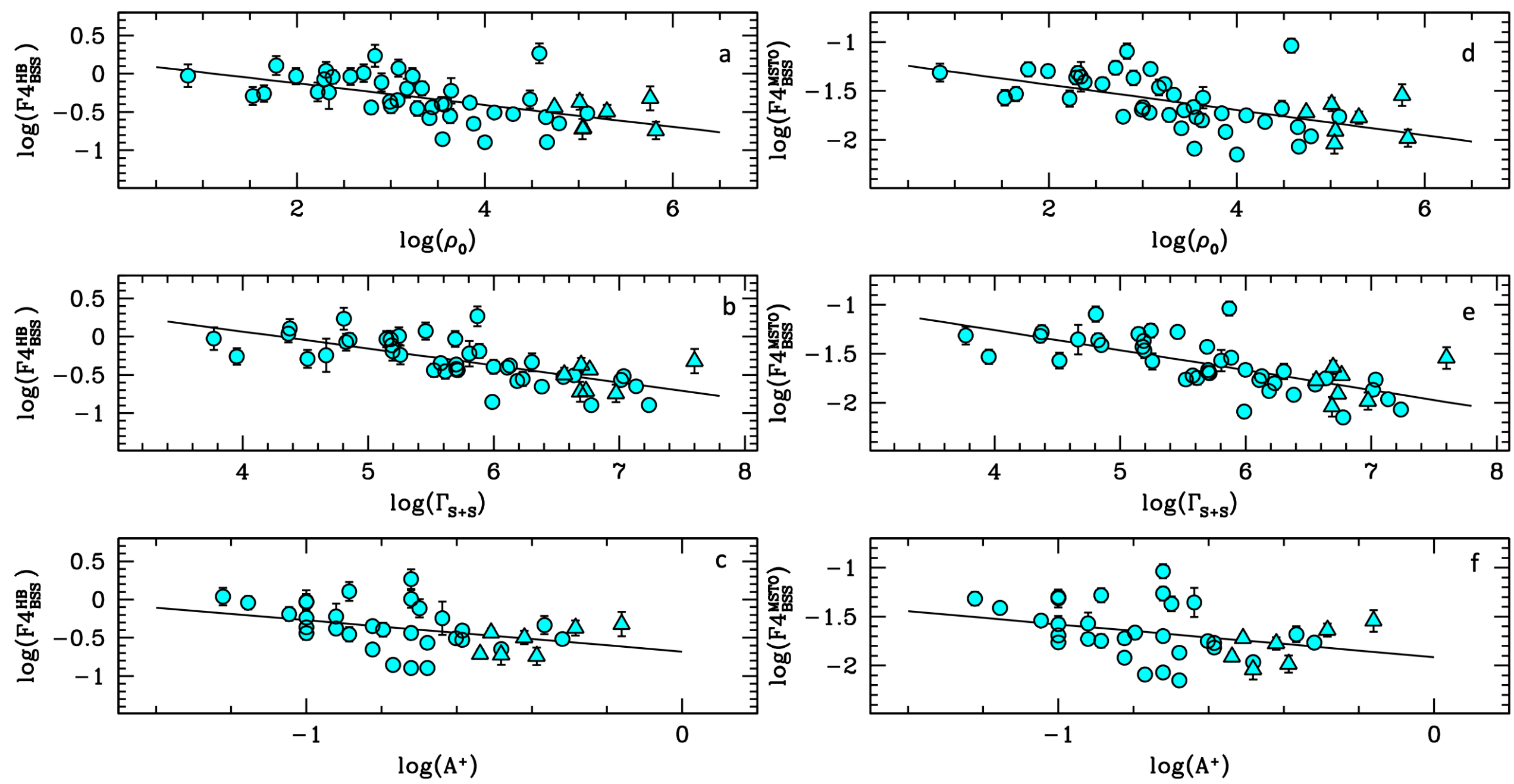

**Supplementary Fig. 3 | Changing the normalization of the blue straggler frequency.** The main relations shown in Figure 2 persist even when the number of blue straggler stars is normalized to the number of horizontal branch stars ($F4^{HB}_{BSS}$, left panels), and to the number of stars in the main sequence turnoff region ($F4^{MSTO}_{BSS}$, right panels). Panels a, b, c, show, respectively, $F4^{HB}_{BSS}$ as a function of the central density ($\rho_0$, in units of $L_\odot/pc^3$), the single-single collisional parameter ($\Gamma_{S+S}$, in arbitrary units), and the dynamical age (parametrized by the $A^+$ parameter, in arbitrary units). Panels d, e, f show, respectively, $F4^{MSTO}_{BSS}$ as a function of the central density ($\rho_0$, in units of $L_\odot/pc^3$), the single-single collisional parameter ($\Gamma_{S+S}$, in arbitrary units), and the dynamical age (parametrized by the $A^+$ parameter, in arbitrary units). In all the panels the best-fit relations to the data are shown as solid black lines. Errors are 1 s.e.m. Source data are provided as a Source Data file.

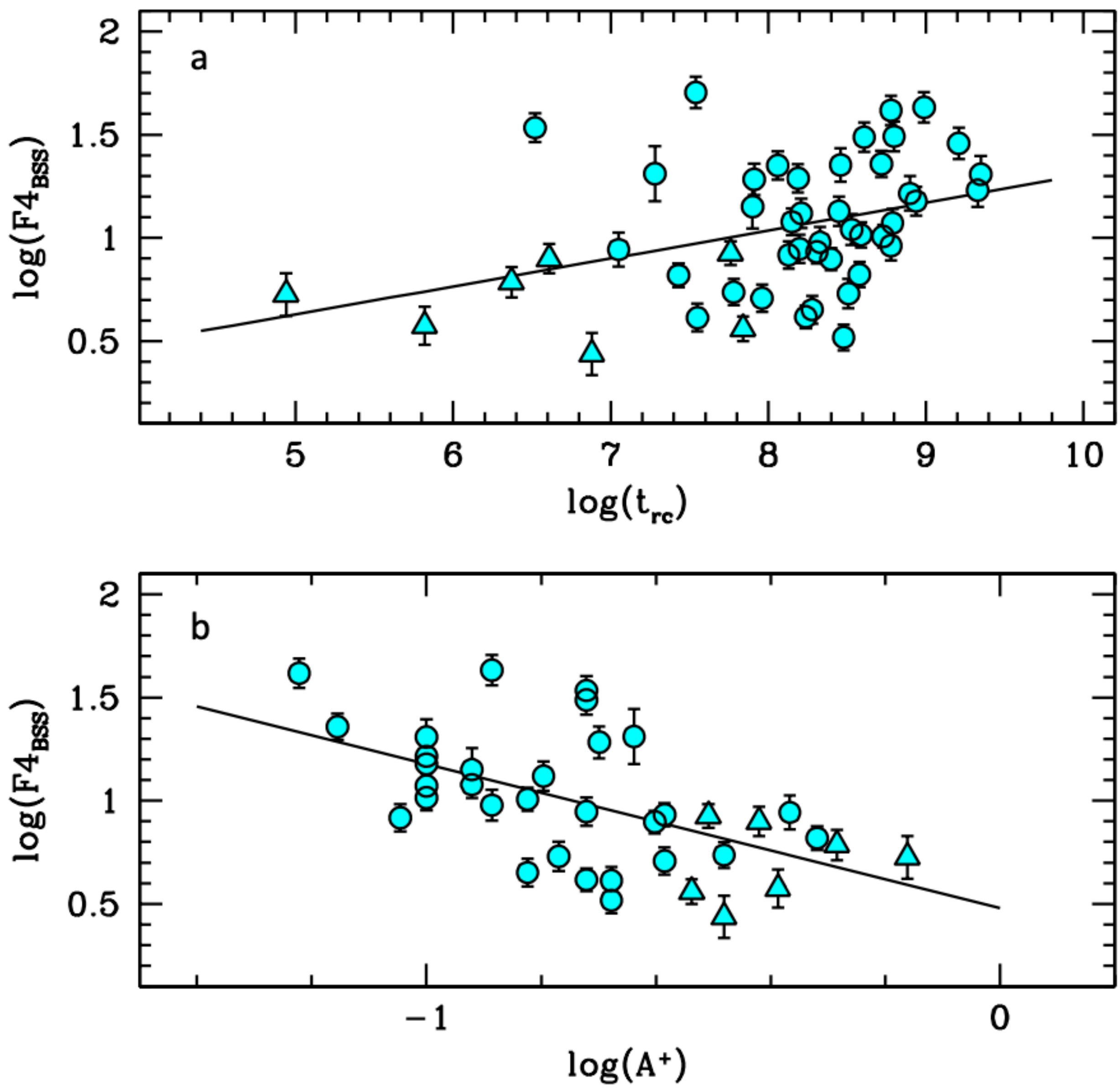

**Supplementary Fig. 4 | Linking the blue straggler frequency with the parent cluster dynamical ageing.** The figure shows the dependence of the blue straggler specific frequency ($F4_{BSS}$, in units of number of stars/$10^4$ $L_{\odot}$) on the parent cluster dynamical ageing quantified by the central relaxation time[19] ($t_{rc}$ in units of years, with dynamically more evolved clusters having lower values of $t_{rc}$; panel a), and by the $A^+$ parameter (in arbitrary units, with dynamically more evolved clusters having larger values of $A^+$; panel b). The best-fit relations to the data are shown as solid black lines. In both cases the blue straggler frequency decreases when the cluster dynamical age increases. However, the correlation is better defined when $A^+$ is used: the Pearson correlation coefficient increases from 0.38 (for the trend with of $t_{rc}$), to 0.56 (for the relation with $A^+$). Errors are 1 s.e.m. Source data are provided as a Source Data file.

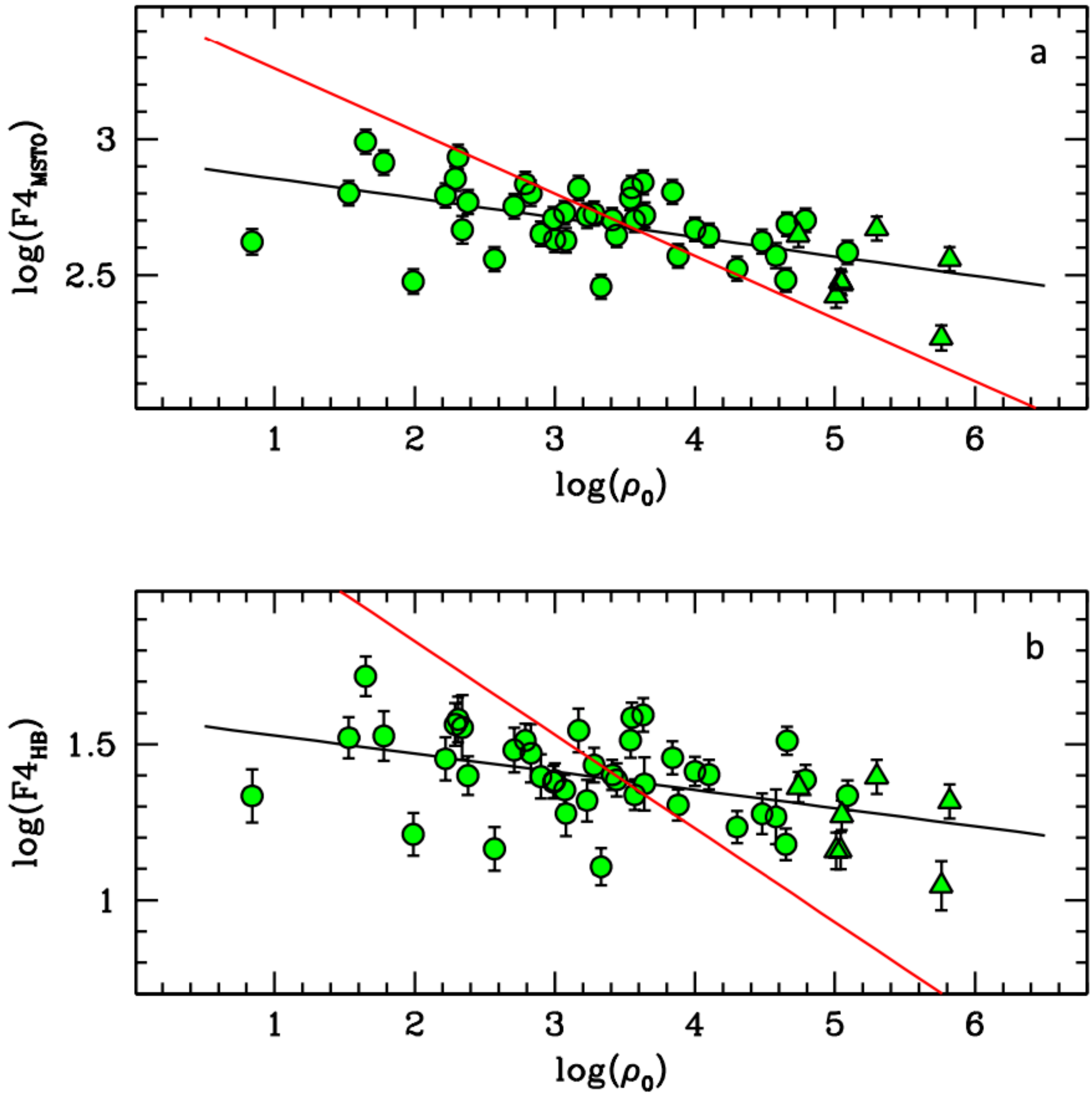

**Supplementary Fig. 5 | Binaries cannot account for the trends observed between normal populations and environment.** The figure illustrates that it is not possible to reproduce the trend (black line in panel a) between the specific frequency of main sequence turnoff stars ($F4_{MSTO}$, in units of number of stars/$10^4$ $L_{\odot}$) and the central cluster density ($\rho_0$, in units of $L_{\odot}/pc^3$), by combining the $f_{bin}$ -$\rho_0$ relation with the $f_{bin}$-$F4_{MSTO}$ one (red line in panel a), with $f_{bin}$ being the cluster binary fraction. The same holds if horizontal branch (HB) stars are considered (see panel b). Errors are 1 s.e.m. Source data are provided as a Source Data file.

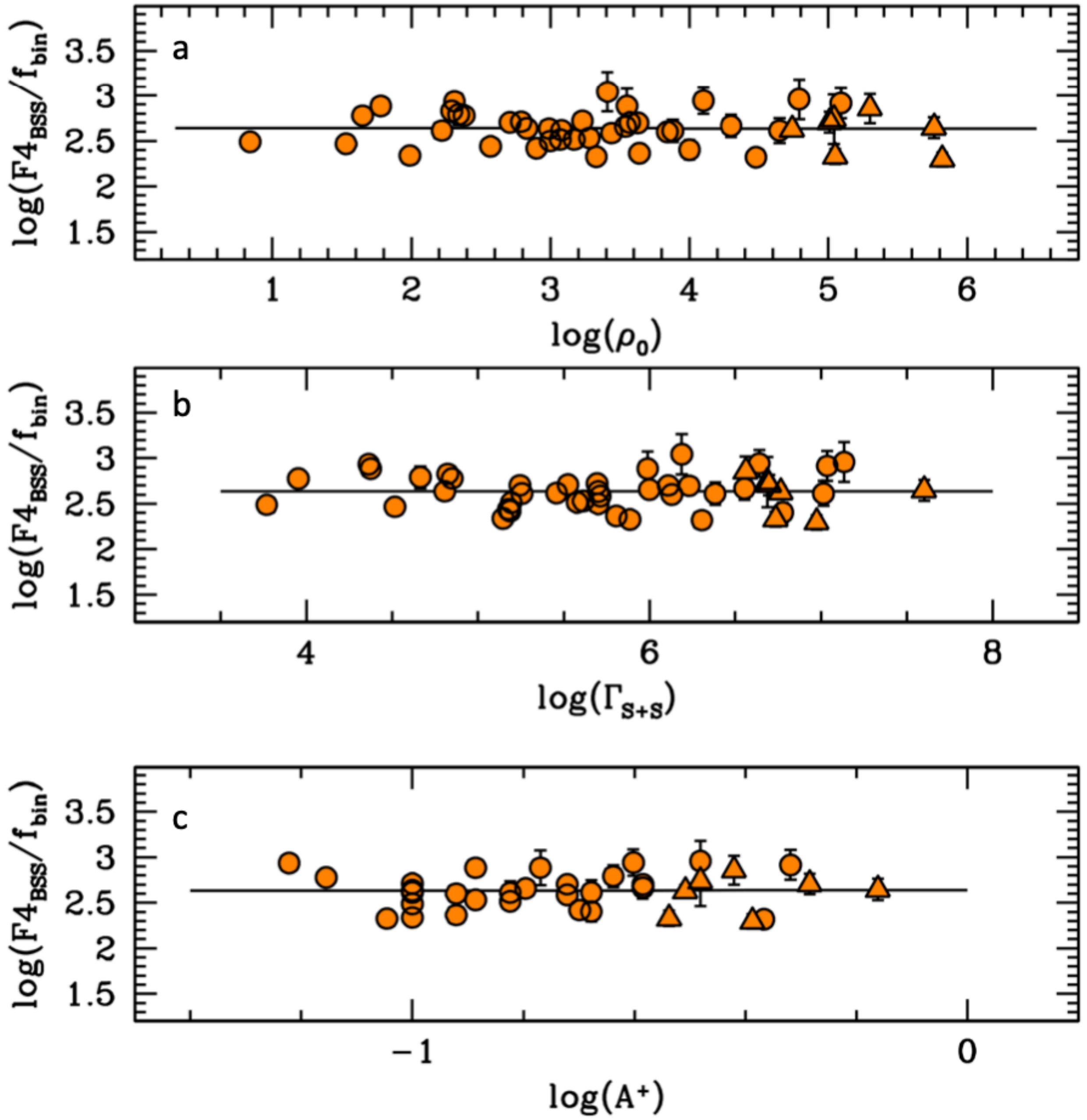

**Supplementary Fig. 6 | The blue straggler specific frequency and the binary fraction scales in the same way.** The figure shows that all the detected trends plotted in Figure 2d,e,f between the blue straggler specific frequency ($F4_{BSS}$, in units of number of stars/$10^4$ $L_\odot$) and the cluster environmental parameters (central density $\rho_0$, collisional parameter $\Gamma_{S+S}$, and dynamical age $A^+$, panels a, b, c, respectively) are completely cleared out when the ratio between $F4_{BSS}$ and the binary fraction ($f_{bin}$) is considered, once more demonstrating that the blue straggler frequency and the binary fraction scale in the same way with those parameters. The central density $\rho_0$ is in units of $L_\odot/pc^3$, the collisional and $A^+$ parameters are in arbitrary units. Errors (1 s.e.m) are calculated by following the standard error propagation law. Source data are provided as a Source Data file.

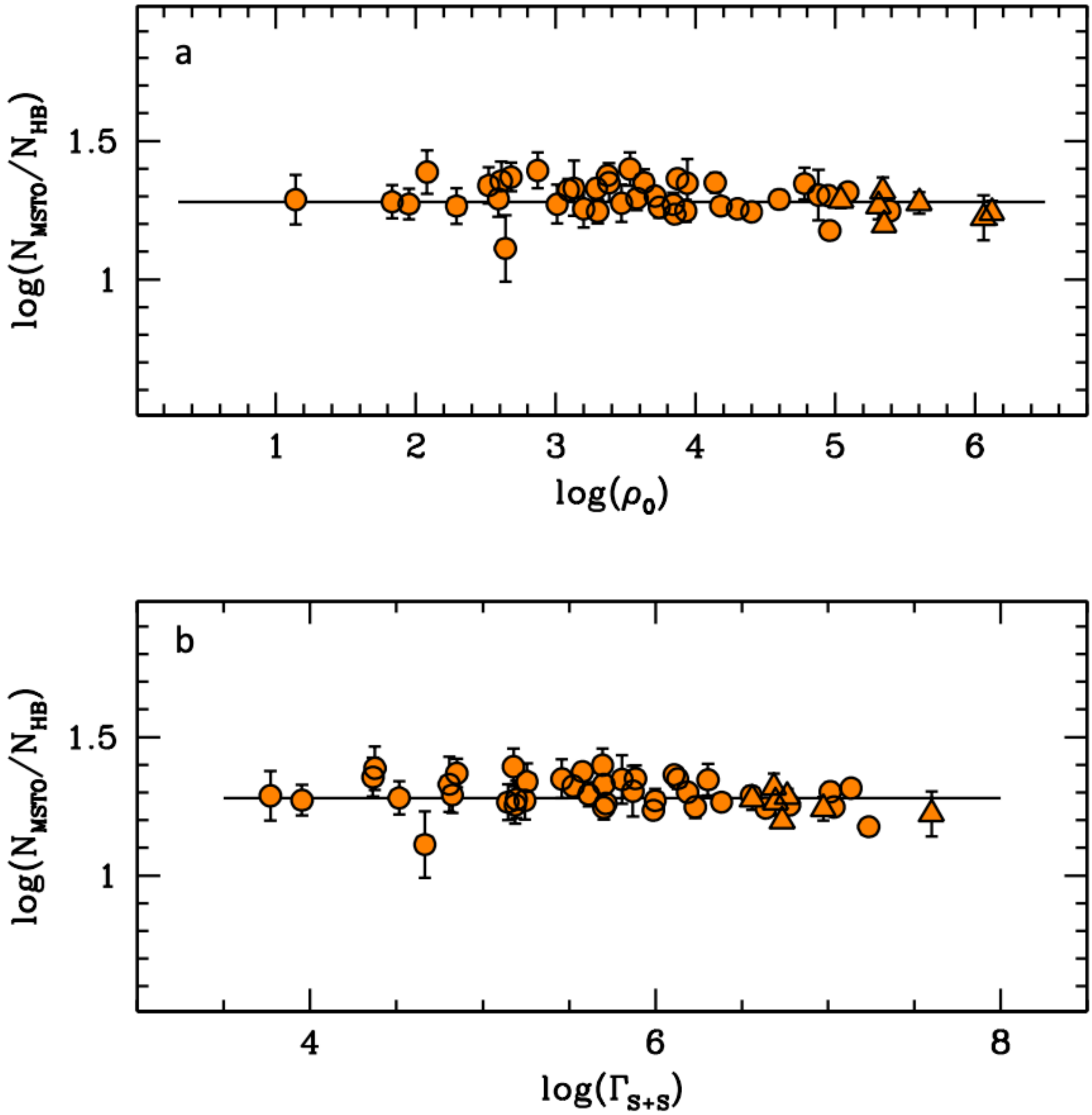

**Supplementary Fig. 7 | Investigating the completeness.** The figure shows that the ratio between the number of stars in the main sequence turnoff (MSTO) region and the number of horizontal branch (HB) stars, as a function of the cluster central density ($\rho_0$ in units of $L_\odot/pc^3$, panel a) and the collisional parameter ($\Gamma_{S+S}$ in arbitrary units, panel b) is constant. The lack of any trend is indeed expected for complete samples of stars. Hence, this confirms that, even a sub-population fainter than blue straggler stars (i.e., MSTO stars) is affected by negligible incompleteness. Errors (1 s.e.m) are calculated by following the standard error propagation law. Source data are provided as a Source Data file.

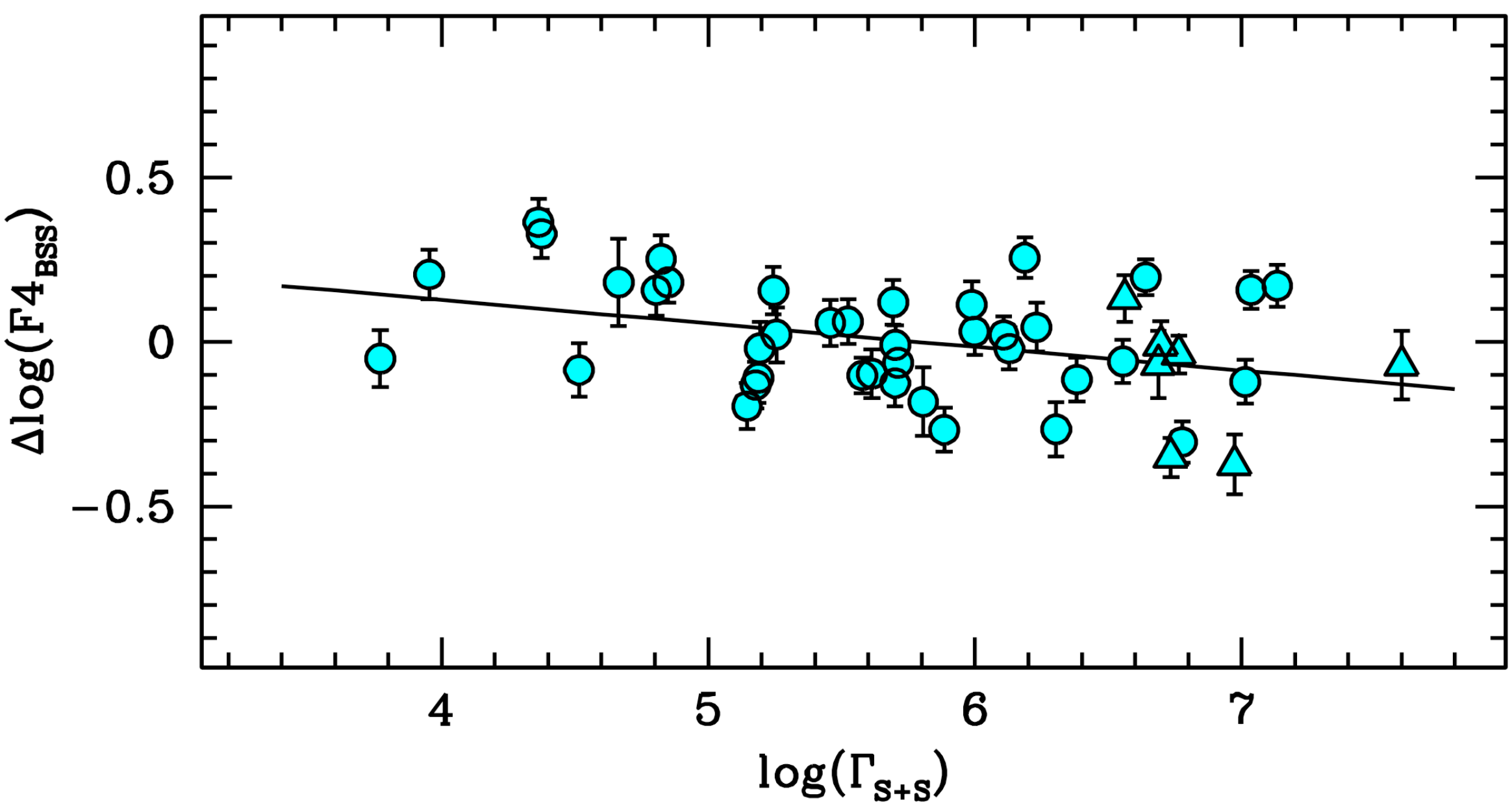

**Supplementary Fig. 8 | Searching for secondary effects.** Trend of the residuals $\Delta\log(F4_{BSS})$ with respect to the best-fit $\log(F4_{BSS})$-$\log(f_{bin})$ relation shown in Figure 4, as a function of the collisional parameter ($\Gamma_{S+S}$ in arbitrary units). Errors are 1 s.e.m. Source data are provided as a Source Data file.

**Supplementary Table 1 | The main best-fit relations.** Values of the slopes and intercepts (with their 1 s.e.m errors) of the best-fit straight lines shown as solid lines in the Main Figures listed in the first column.

| Figure | Best fit relations | Pearson Correlation Coefficient |
|---|---|---|
| Figure 2a | $\log(F4_{BSS}) = 0.23 (\pm 0.04) \times M_V + 2.79 (\pm 0.27)$ | 0.69 |
| Figure 2b | $\log(F4_{BSS}) = -0.09 (\pm 0.01) \times \sigma_0 + 1.58 (\pm 0.07)$ | –0.78 |
| Figure 2c | $\log(F4_{BSS}) = 0.13 (\pm 0.02) \times \mu_{V,0} - 1.18 (\pm 0.26)$ | 0.78 |
| Figure 2d | $\log(F4_{BSS}) = -0.20 (\pm 0.03) \times \log(\rho_0) + 1.75 (\pm 0.10)$ | –0.72 |
| Figure 2e | $\log(F4_{BSS}) = -0.30 (\pm 0.03) \times \log(\Gamma_{S+S}) + 2.77 (\pm 0.17)$ | –0.84 |
| Figure 2f | $\log(F4_{BSS}) = -0.68 (\pm 0.15) \times \log(A^+) + 0.49 (\pm 0.12)$ | –0.60 |
| Figure 3a | $\log(f_{bin}) = 0.24 (\pm 0.045) \times M_V + 0.23 (\pm 0.34)$ | 0.63 |
| Figure 3b | $\log(f_{bin}) = -0.10 (\pm 0.01) \times \sigma_0 - 0.97 (\pm 0.08)$ | –0.79 |
| Figure 3c | $\log(f_{bin}) = 0.13 (\pm 0.02) \times \mu_{V,0} - 3.91 (\pm 0.30)$ | 0.76 |
| Figure 3d | $\log(f_{bin}) = -0.21 (\pm 0.03) \times \log(\rho_0) - 0.87 (\pm 0.12)$ | –0.70 |
| Figure 3e | $\log(f_{bin}) = -0.30 (\pm 0.04) \times \log(\Gamma_{S+S}) + 0.13 (\pm 0.21)$ | –0.78 |
| Figure 3f | $\log(f_{bin}) = -0.68 (\pm 0.16) \times \log(A^+) - 2.15 (\pm 0.12)$ | –0.60 |
| Figure 4 | $\log(f_{bin}) = 0.91 (\pm 0.09) \times \log(F4_{BSS}) - 2.55 (\pm 0.10)$ | 0.83 |